\documentclass[aip,cha,reprint]{revtex4-2}

\makeatletter
\AtBeginDocument{\let\selectlanguage\@gobble}
\makeatother
\usepackage{amsmath,amssymb}
\usepackage{mathrsfs}
\usepackage{bm}
\usepackage{graphicx}
\usepackage{newfloat}
\DeclareFloatingEnvironment[name=Algorithm, placement=t, fileext=loa]{algorithm}
\usepackage{algpseudocode}
\usepackage[hidelinks]{hyperref}

\algrenewcommand\algorithmicrequire{\textbf{Input:}}
\algrenewcommand\algorithmicensure{\textbf{Output:}}

\begin{document}

\title{Slow-spectrum preshaping for reservoir computing: \\
Attractor reconstruction in attracting submanifolds under partial observation}

\author{Satoshi Oishi}
\email{satoshi.oishi@ist.osaka-u.ac.jp}
\affiliation{
Graduate School of Information Science and Technology, The University of Osaka,
1-5 Yamadaoka, Suita, Osaka 565-0871, Japan}
\author{Hiroshi Yamashita}
\affiliation{
International Research Center for Neurointelligence, The University of Tokyo,
7-3-1 Hongo, Bunkyo-ku, Tokyo 113-0033, Japan}
\affiliation{
Graduate School of Information Science and Technology, The University of Osaka,
1-5 Yamadaoka, Suita, Osaka 565-0871, Japan}
\author{Hideyuki Suzuki}
\author{Sho Shirasaka}
\affiliation{
Graduate School of Information Science and Technology, The University of Osaka,
1-5 Yamadaoka, Suita, Osaka 565-0871, Japan}
\affiliation{
International Research Center for Neurointelligence, The University of Tokyo,
7-3-1 Hongo, Bunkyo-ku, Tokyo 113-0033, Japan}

\date{\today}

\begin{abstract}
Data-driven reproduction of chaotic dynamics under partial observation remains a challenge despite its practical importance. Reservoir computing (RC) and other data-driven approaches often succeed in short-term prediction but are sensitive to hyperparameters and frequently fail to reproduce the long-term statistical properties of the system. As recently shown, a major cause of this failure is spurious slow modes induced by training, which render the reconstructed invariant set transversally unstable in the representation space. To address this, a design principle called Attractor Reconstruction in Attracting Submanifolds (ARAS) was proposed, which requires the reconstruction to lie in a low-dimensional, transversally attracting submanifold. Under full observation, ARAS was realized by an input-layer design that limits the number of slow modes perturbed during training and anchors the remaining modes to stay fast, thereby suppressing spurious slow modes. However, under partial observation, this design alone fails because the submanifold designed to be attracting cannot retain the observation history needed to unfold the observed data into a faithful reconstruction. In this study, we propose \textit{slow-spectrum preshaping}, which introduces a slow spectrum into the reservoir prior to feeding input data. The induced slow modes then retain a memory of past observations within the submanifold, while the input-layer design makes the submanifold transversally attracting, so that ARAS is realized under partial observation. Through numerical experiments on various chaotic systems, we show that our approach enables reliable short-term prediction and long-term reproduction of chaos over a wide range of hyperparameters without \textit{a posteriori} tuning.
\end{abstract}

\maketitle

\begin{quotation}
In real-world applications, time-series forecasting is often performed under partial observation, and various machine-learning methods have been developed to tackle this challenge. Although these models typically forecast the observed signal well for a short time, they frequently fail to reproduce the system's long-term behavior once they run autonomously. A major cause of such failures is that the target attractor is reconstructed in a dynamically unstable region of the model's internal state space. To address this, a design principle called Attractor Reconstruction in Attracting Submanifolds (ARAS) was recently proposed, which requires the reconstructed attractor set to lie in a region that attracts nearby trajectories. While a previous method successfully achieves ARAS under full observation by limiting the number of potentially unstable modes, it struggles when only partial measurements are available. In this paper, we propose a slow-spectrum preshaping technique for reservoir computing, a type of recurrent neural network well-suited for physical AI and edge AI, which intentionally introduces slow dynamical modes to retain the history of past observations. This approach successfully realizes ARAS even under partial observation, reproducing the long-term behavior of chaotic systems far more reliably than standard designs, without relying on extensive hyperparameter tuning. The idea of ARAS and slow-spectrum preshaping may also help other data-driven methods that embed a system's dynamics in a high-dimensional space.
\end{quotation}

\section{Introduction}\label{sec:introduction}

Data-driven forecasting and modeling of nonlinear chaotic systems are important for understanding and controlling complex phenomena across many scientific fields~\cite{Brunton2022-bj}. In most real applications, however, only a few state variables can be measured, and reconstructing the underlying dynamics from such partial observations remains difficult~\cite{Kantz2004-wo,Ozalp2023-do}.

In the analysis of partially observed chaotic time series, attractor reconstruction~\cite{Packard1980-ay}---mapping the observed sequence into a high-dimensional space so that the geometric structure of the original attractor is recovered---has played a central role. Embedding theorems~\cite{Takens1981-mw,Sauer1991-ew} guarantee that an attractor can generically be reconstructed from time-shifted copies of an observed signal. The resulting delay-coordinate maps underlie a wide range of time-series analysis methods such as the Hankel alternative view of Koopman (HAVOK) framework~\cite{Brunton2017-xi}, convergent cross mapping (CCM)~\cite{Sugihara2012-in}, and sliding-window methods in topological data analysis (TDA)~\cite{Perea2015-xp}. Other than delay coordinates, the idea of lifting observations into a high-dimensional space is now ubiquitous in machine learning and data-driven modeling such as extended dynamic mode decomposition (EDMD)~\cite{Williams2015-jg}, sparse identification of nonlinear dynamics (SINDy)~\cite{Brunton2016-gc}, and next-generation reservoir computing (NGRC)~\cite{Gauthier2021-pz, Bollt2021-mj}. These approaches map observed signals into high-dimensional spaces using delay and/or nonlinear features to capture dynamical structure that is difficult to extract in the original observation space. Other recent machine-learning approaches that leverage high-dimensional lifting include kernel methods, which linearize nonlinear data by lifting it into infinite-dimensional feature spaces~\cite{Scholkopf2002-wd,Kawahara2016-ai}, modern state-space models (SSMs), which embed the history of the input sequence into a high-dimensional recurrent state~\cite{Gu2021-uh, Gu2023-xy}, and deep autoencoder-based models, which discover intrinsic coordinates that facilitate the representation of governing dynamics~\cite{Lusch2018-yn, Champion2019-nc, Bakarji2023-qa}.

Among these approaches, reservoir computing (RC)~\cite{Jaeger2001-vs, Maass2002-me, Jaeger2007-py} has long attracted attention for the prediction of chaotic dynamics. In RC, the high-dimensional latent dynamics are generated by a fixed, often randomly initialized reservoir driven by a fixed input mapping. This framework is highly suitable for physical AI and edge AI because the fixed reservoir can be implemented using high-speed, energy-efficient physical devices~\cite{Tanaka2019-vn} and only the output weights are trained by linear regression. Despite this simple architecture, RC succeeds in short-term prediction of chaotic time series~\cite{Pathak2017-id}. This success is attributed to the target attractor being embedded in the high-dimensional reservoir state space in a manner analogous to delay coordinates~\cite{Lu2018-vk,Grigoryeva2021-mg}. If the reconstructed invariant set corresponding to the target attractor is transversally stable, the RC can emulate the long-term dynamics of the target system while preserving its qualitative properties.

\begin{figure}[htbp]
  \centering
  \includegraphics[width=\columnwidth]{figures/Fig1}
\caption{Schematic of attractor reconstruction in the reservoir state space. (a) In standard RC, the submanifold where reconstruction occurs cannot be explicitly controlled, and it is not guaranteed to be attracting. 
(b) In the RC constructed using the deterministic input-layer design of Ref.~\onlinecite{Oishi2026-fz} under full observation, the attractor can be reconstructed in a transversally attracting submanifold (ARAS). (c) In the RC constructed using the same design under partial observation, the reconstructed attractor collapses to a low-dimensional structure during closed-loop operation. Slow-spectrum-preshaped RC unfolds the history of partially observed data in the slow spectral submanifold. By applying the framework of Ref.~\onlinecite{Oishi2026-fz}, this slow submanifold becomes transversally attracting, which stabilizes the reconstructed attractor and enables robust reproduction of the dynamics.}
  \label{fig:concept}
\end{figure}

However, the performance of RC is known to depend strongly on the random initialization of the input mapping and the latent recurrent dynamics and on numerous hyperparameters~\cite{Platt2022-es}. In a standard RC, there is no explicit control over where the target attractor is reconstructed in the vast representation space, and if it is embedded in a transversally unstable region, the closed-loop trajectory may deviate from the reconstructed invariant set [Fig.~\ref{fig:concept}(a)]. Note that such failures of autonomous prediction occur even when the attractor is correctly reconstructed, that is, an embedding is achieved, inside the reservoir. An embedding alone does not guarantee the reproduction of the dynamics because the embedded invariant set is not necessarily transversally stable~\cite{Lu2018-vk,Hart2024-jx}. No \textit{a priori} construction had hitherto been available that guarantees the transverse stability of the reconstructed invariant set corresponding to the target attractor, and practitioners had instead relied on \textit{a posteriori} hyperparameter tuning guided by performance metrics~\cite{Yperman2016-xd, Bala2018-zq, Racca2021-xx} or by the reproduction of dynamical invariants~\cite{Platt2023-lh}. To address this, a design principle called Attractor Reconstruction in Attracting Submanifolds (ARAS) was recently proposed~\cite{Oishi2026-fz}, which requires the reconstructed invariant set to be placed in a transversally attracting submanifold [Fig.~\ref{fig:concept}(b)]. For echo state networks (ESNs)~\cite{Jaeger2001-vs}, a widely used form of RC implemented as recurrent neural networks (RNNs), this principle was realized by an \textit{a priori}, task-agnostic deterministic input-layer design that constructs the input layer from a low-dimensional invariant subspace of the reservoir transition matrix~\cite{Oishi2026-fz}. This design limits the number of emerging slow modes in closed-loop (autonomous) operation and thus suppresses spurious slow modes, which can be transversally unstable. In this way, it achieves robust reproduction of dynamical systems under full observation.

We found, however, that an RC with this prescribed input layer fails severely when applied directly under partial observation (see also Sec.~\ref{sec:deterministic-win-limitations}). Standard RC also frequently fails to predict or reproduce dynamics in this setting~\cite{Mahata2023-dw}. These failures under partial observation undermine the overall reliability of RC as a modeling tool. Through numerical experiments, we confirm that a major cause of these failures is that the submanifold designed to be attracting cannot retain the observation history, so that the attractor reconstructed within it collapses into an inappropriately low-dimensional structure. To robustly realize ARAS for RC under partial observation, we propose \textit{slow-spectrum preshaping}: a small number of slow eigenvalues are placed in the reservoir transition matrix in advance, and the input layer is then designed from a \textit{slow spectral subspace}, an invariant subspace associated with these slow eigenvalues. This preshaping introduces slow modes associated with these eigenvalues, which form the \textit{slow (spectral) submanifold} \cite{Haller2016-oj} and retain a memory of past inputs. The geometric structure of the attractor is then reconstructed within this submanifold by the same principle as delay-coordinate embedding. Furthermore, by the same mechanism as in the ARAS realization under full observation~\cite{Oishi2026-fz}, this slow submanifold becomes transversally attracting, so the reconstructed invariant set is kept transversally stable and the dynamics can be reproduced robustly [Fig.~\ref{fig:concept}(c)].

This paper is organized as follows. Section~\ref{sec:rc-fundamentals} introduces standard RC. Section~\ref{sec:standard-rc-limitations} demonstrates, through preliminary experiments, the difficulty of chaotic prediction under partial observation, and Sec.~\ref{sec:deterministic-win-limitations} revisits our deterministic input-layer design~\cite{Oishi2026-fz}, which functions as an RC implementation of ARAS under full observation but fails on its own under partial observation. To overcome this limitation, Sec.~\ref{sec:proposed-methodology} presents the slow-spectrum preshaping framework and explains its underlying mechanism. Section~\ref{sec:comprehensive-evaluation} provides a comprehensive numerical evaluation. Section~\ref{sec:esp-verification} examines the concern that introducing slow eigenvalues may destroy the contractivity of the reservoir evolution (the echo state property) and verifies, through conditional Lyapunov exponents, that the contractivity in the sense of generalized synchronization is maintained. Section~\ref{sec:attractor-reconstruction-performance} demonstrates, by estimating the Lyapunov spectrum, that the proposed method reproduces key dynamical properties of the target system, and Sec.~\ref{sec:parameter-robustness-evaluation} evaluates the prediction performance over hyperparameter space, showing that the proposed method provides stable modeling over a much wider parameter range than standard RC. Finally, Sec.~\ref{sec:conclusion-and-discussion} discusses the limitations of the proposed method and future directions, and concludes the paper.

\section{Reservoir computing and problem formulation}
\label{sec:rc-fundamentals}
RC~\cite{Jaeger2001-vs,Jaeger2007-kc,Tanaka2019-vn} is a machine-learning framework in which a nonadaptive complex dynamical system is driven by fixed input channels and only the output weights are trained, for example by linear regression. This section reviews the standard RC procedure for time-series prediction based on ESNs, an RNN implementation of RC, and formulates the problem addressed in this paper.

We formulate the target dynamical system to be predicted and reproduced as
\begin{equation}
\bm{x}_{t+1} = F(\bm{x}_t),
\label{eq:target-system}
\end{equation}
where $\bm{x}_t \in \mathscr{M}$ is the state, $\mathscr{M}$ is the state space, and $F: \mathscr{M} \to \mathscr{M}$ is the time-evolution map (or flow). Observations are obtained through an observation function $h: \mathscr{M} \to \mathbb{R}^D$ as $\bm{u}_t = h(\bm{x}_t)$.

The observation sequence $\{\bm{u}_t\}$ is fed into the RC, whose internal state $\bm{r}_t \in \mathbb{R}^N$ is updated recursively by the leaky-integrator rule
\begin{equation}
\bm{r}_{t+1} = f(\bm{r}_{t}, \bm{u}_t)
= (1-\alpha) \bm{r}_{t}
+ \alpha\, \bm{\tau}(\mathbf{A} \bm{r}_{t}
+ \mathbf{W}_{\mathrm{in}} \bm{u}_t + \bm{b}),
\label{eq:rc-update}
\end{equation}
where $\mathbf{A} \in \mathbb{R}^{N \times N}$ is the reservoir transition matrix, $\mathbf{W}_{\mathrm{in}} \in \mathbb{R}^{N \times D}$ is the input-layer matrix, and $\bm{b} \in \mathbb{R}^N$ is a bias vector. All three are randomly initialized and fixed thereafter. The parameter $\alpha \in (0, 1]$ is the leaking rate, $N$ is the number of reservoir nodes, and $\bm{\tau}$ denotes the activation function. Regarding the dynamical system~\eqref{eq:target-system} as driving Eq.~\eqref{eq:rc-update}, we can formulate the combined system as the skew-product system
\begin{equation}
\left\{
\begin{aligned}
\bm{x}_{t+1} &= F(\bm{x}_t), \\
\bm{r}_{t+1} &= f(\bm{r}_{t}, h(\bm{x}_t)).
\end{aligned}
\right. \label{eq:skew-product}
\end{equation}

For reliable information processing in RC, the reservoir should respond reproducibly to the driving signal, independently of its initial state. Even for the same input sequence $\{\bm{u}_t\}$, reservoir state sequences $\{\bm{r}_t\}$ generated by Eq.~\eqref{eq:rc-update} from different initial states $\bm{r}_0$ may differ. If, however, the update map $f$ is a contraction with respect to the internal state, then any two trajectories started from different initial states $\bm{r}_0 \neq \bm{r}'_0$ and driven by the same input sequence converge to each other asymptotically, i.e., $\|\bm{r}_t - \bm{r}'_t\| \to 0$ as $t \to \infty$. This loss of dependence on the initial condition is known as generalized synchronization (GS) in drive--response systems~\cite{Rulkov1995-tq,Pyragas1996-kv,Lu2018-vk,Suetani2026-hg}. In the skew-product picture of Eq.~\eqref{eq:skew-product}, there then exists, in the limit $t \to \infty$, a  map $\bm{r}_t = \Phi(\bm{x}_t)$ from the target state to the reservoir state. A related property in the RC literature is the echo state property (ESP)~\cite{Jaeger2001-vs, Manjunath2013-bj}, which guarantees that for any input sequence belonging to a specified subset of bounded sequences---not only the specific signal driving the reservoir---the reservoir state asymptotically becomes a unique function of the input history, independent of the initial condition of the reservoir. 

Throughout this paper, unless otherwise stated, we use the elementwise hyperbolic tangent $\bm{\tau} = \bm{\tanh}$ as the activation function, set $\mathbf{A}$ to be a real symmetric matrix, and scale it so that the spectral radius of the effective reservoir transition matrix $\mathbf{A}_{\mathrm{eff}} := (1-\alpha)\mathbf{I} + \alpha \mathbf{A}$ (which coincides with its maximum singular value since $\mathbf{A}$ is symmetric) is strictly less than one. For the unforced system ($\bm{u}_t \equiv \bm{0}$), the Jacobian matrix evaluated at the equilibrium point $\bm{r}_*$ is $\mathbf{J} = (1-\alpha)\mathbf{I} + \alpha \mathbf{D} \mathbf{A}$, where $\mathbf{D} = \mathrm{diag}(\mathbf{1} - \bm{\tanh}^2(\mathbf{A}\bm{r}_* + \bm{b}))$. When the maximum singular value of $\mathbf{J}$ is strictly less than one and the bias $\bm{b}$ is small, the equilibrium point that reduces to the origin for $\bm{b}=\bm{0}$ lies close to the origin, so that $\bm{r}_* \approx \bm{0}$ and hence $\mathbf{D} \approx \mathbf{I}$. Accordingly, the linearized dynamics around the equilibrium are approximately governed by $\mathbf{A}_{\mathrm{eff}}$. Thus, the assumptions above make $f$ a uniform contraction~\cite{Jaeger2001-vs} provided that $\bm{b}$ is sufficiently small.

In practice, to remove the dependence on the initial state, the reservoir is first run for a sufficiently long warm-up period of $T_\mathrm{w}$ steps. After the warm-up, a reservoir state sequence of length $T$, $\mathbf{R}_T := \{\bm{r}_t\}_{t=T_\mathrm{w}}^{T_\mathrm{w}+T-1}$, is collected for training. This process of driving the reservoir with the input sequence is called the \textit{listening phase}. During this phase, as with delay-coordinate reconstruction grounded in Takens' embedding theorem~\cite{Takens1981-mw}, a sufficiently large RC is expected to generically provide an embedding of the target invariant set $\mathcal{I} \subset \mathscr{M}$ (i.e., $\Phi$ is a diffeomorphism from $\mathcal{I}$ onto its image) even under partial observation~\cite{Pathak2017-id, Lu2018-vk}. This expectation, referred to as the ESN Embedding Conjecture~\cite{Hart2020-kx}, has been rigorously proven when the activation function $\bm{\tau}$ is linear~\cite{Grigoryeva2021-mg, Grigoryeva2023-cw}.

We next describe the \textit{training phase}, in which the output layer is optimized. Using the reservoir state sequence $\mathbf{R}_T$ obtained in the listening phase, the output weight matrix $\mathbf{W}_{\mathrm{out}} \in \mathbb{R}^{D \times N}$ is determined so that the readout approximates the next-step observation. The standard practice is to minimize the error in a least-squares sense while applying regularization. A common choice is Tikhonov regularization (ridge regression)~\cite{Bishop2006-iz}, which minimizes
\begin{equation}
    \mathbf{W}_{\mathrm{out}} = \operatorname*{argmin}_{\mathbf{W}}
    \left( \sum_{t=T_\mathrm{w}}^{T_\mathrm{w}+T-1}
    \| \bm{u}_t - \mathbf{W} \bm{r}_t \|^2
    + \beta \| \mathbf{W} \|^2 \right),
    \label{eq:ridge-regression}
\end{equation}
where $\beta \geq 0$ is the regularization parameter.

After training, the RC can be operated as an autonomous prediction system in the \textit{prediction phase} (closed-loop prediction). In this phase, the RC's own output $\hat{\bm{u}}_t = \mathbf{W}_{\mathrm{out}} \bm{r}_{t}$ is fed back recursively as the current input in place of the actual observation $\bm{u}_t$, so that the reservoir update becomes autonomous:
\begin{align}
  \bm{r}_{t+1}
    &= \hat{f}(\bm{r}_{t})
    := (1-\alpha) \bm{r}_{t} + \alpha\, \bm{\tau}(\mathbf{A} \bm{r}_{t}
    + \mathbf{W}_{\mathrm{in}} \mathbf{W}_{\mathrm{out}} \bm{r}_{t}
    + \bm{b}) \nonumber \\
    &= (1-\alpha) \bm{r}_{t}
    + \alpha\, \bm{\tau}(\mathbf{W}_{\mathrm{cl}} \bm{r}_{t} + \bm{b}),
    \label{eq:closed-loop-dynamics}
\end{align}
where the closed-loop matrix is defined as $\mathbf{W}_{\mathrm{cl}} := \mathbf{A} + \mathbf{W}_{\mathrm{in}} \mathbf{W}_{\mathrm{out}}$. Through this autonomous closed loop, the RC continues to generate future trajectories without being supplied with observation data. An appropriately constructed and trained RC not only attains accurate short-term trajectory prediction but is also known to reproduce long-term properties of the target system, such as the statistical properties of its attractor, including the Lyapunov spectrum~\cite{Pathak2017-id, Lu2018-vk}.

As noted above, under suitable conditions the invariant set of the target system can be embedded into the reservoir state space even under partial observation of the state~\cite{Grigoryeva2023-cw}. However, the fact that the topology is preserved during the listening phase does not guarantee successful long-term reproduction of the target attractor in closed-loop prediction. The generated trajectories often diverge or collapse onto behavior that does not reflect the target attractor if the reconstructed invariant set $\Phi(\mathcal{I})$ is transversally unstable under closed-loop prediction (i.e., if there exists an unstable direction in the reservoir state space transverse to $\Phi(\mathcal{I})$).

Transverse Lyapunov exponents (TLEs), also called conditional Lyapunov exponents (CLEs)~\cite{Fujisaka1983-hu,Pecora2015-gy}, quantify this linear stability by regarding the target system as driving a response system (in this case, the reservoir). The TLEs are obtained by removing the Lyapunov exponents of the target system from the Lyapunov exponents computed on the reconstructed invariant set $\Phi(\mathcal{I})$. Accordingly, the variational equation for the TLE computation is governed by the Jacobian matrix $\mathbf{J}_t$ derived from the closed-loop update rule~\eqref{eq:closed-loop-dynamics},\footnote{The final equality follows from Eq.~\eqref{eq:closed-loop-dynamics}, which allows the $\bm{\tanh}$ term to be expressed directly in terms of the stored state sequence. This form is convenient for numerical computation of the Lyapunov exponents.}
\begin{align}
    \mathbf{J}_t &= \left. \frac{\partial \hat{f}(\bm{r})}{\partial \bm{r}} \right|_{\bm{r} = \bm{r}_t \in \Phi(\mathcal{I})} \nonumber \\
    &= (1-\alpha)\mathbf{I}
    + \alpha\, \left. \frac{\partial}{\partial \bm{r}}
      \bm{\tanh}(\mathbf{W}_{\mathrm{cl}} \bm{r} + \bm{b}) \right|_{\bm{r} = \bm{r}_t } \nonumber \\
    &= (1-\alpha)\mathbf{I} + \alpha\,
      \mathrm{diag}\!\left(1 - \bm{\tanh}^2(\mathbf{W}_{\mathrm{cl}} \bm{r}_t
      + \bm{b})\right) \mathbf{W}_{\mathrm{cl}} \nonumber \\
    &= (1-\alpha)\mathbf{I} + \alpha\,
      \mathrm{diag}\!\left(1 - \tfrac{1}{\alpha^2}
      \left(\bm{r}_{t+1} - (1-\alpha)\bm{r}_t\right)^2 \right)
      \mathbf{W}_{\mathrm{cl}}.
    \label{eq:jacobian}
\end{align}

If at least one TLE is positive, the invariant set reconstructed during the listening phase becomes transversally unstable, and trajectories diverge from it under closed-loop prediction. Stable reproduction of the target attractor requires all TLEs to be negative, but no construction is currently available that satisfies this condition \textit{a priori} in standard RC, and in practice one must rely on heuristic hyperparameter search. Recent studies have reported empirically that the TLEs tend to become positive when the spectral radius of the reservoir transition matrix $\mathbf{A}$ is set very large~\cite{Hart2024-jx}, or when the eigenvalue distribution of the effective closed-loop matrix $\mathbf{W}_{\mathrm{cl,eff}} := (1-\alpha)\mathbf{I} + \alpha \mathbf{W}_{\mathrm{cl}}$ contains an excess number of spurious slow eigenvalues compared to the dimension of the target attractor~\cite{Oishi2026-fz}. Note that for small bias $\bm{b}$, $\mathbf{W}_{\mathrm{cl,eff}}$ approximately governs the linearized dynamics around the equilibrium point that reduces to the origin for $\bm{b}=\bm{0}$.

\subsection{Limitations of standard RC under partial observation}
\label{sec:standard-rc-limitations}

\begin{figure*}[htbp]
  \centering
  \includegraphics[width=\textwidth]{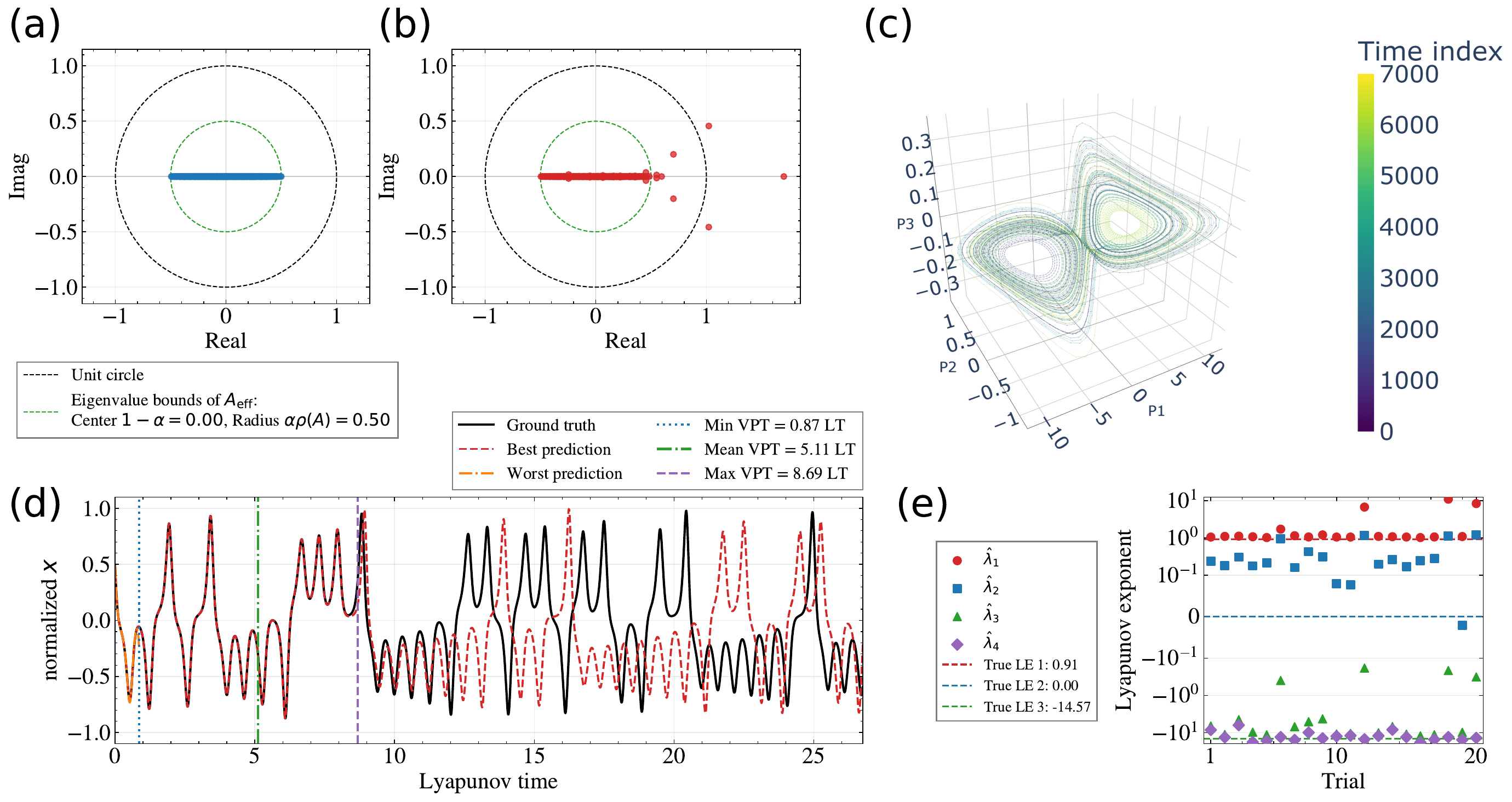}
\caption{Reproduction of the Lorenz-63 system using standard RC. (a) Eigenvalue distribution of $\mathbf{A}_{\mathrm{eff}}$. (b) Eigenvalue distribution of $\mathbf{W}_{\mathrm{cl,eff}}$. (c) Listening-phase reservoir states projected onto their first three principal components. The 3D plot clearly shows the two-lobe structure. (d) Short-term predictions for the realizations with the longest and shortest valid prediction times (VPTs) among 20 independent realizations; the signals are normalized to $[-1, 1]$. Vertical dashed lines indicate the minimum, the mean, and the maximum VPT. (e) Lyapunov spectrum of the closed-loop RC, plotted on a symmetric-log scale that is linear within $[-0.1, 0.1]$. The ground-truth values of the Lorenz attractor are indicated by horizontal dashed lines.}
  \label{fig:standard-rc-results}
\end{figure*}

To illustrate the limitations of standard RC, this subsection reports a preliminary experiment in which only the $x$ variable of the Lorenz-63 system~\cite{Lorenz1963-kf} (see Appendix~\ref{sec:lorenz-equation} for the governing equations and the data-generation procedure) is observable ($D = 1$). The detailed experimental conditions and parameters are identical to those of Sec.~\ref{sec:proposed-methodology} and are summarized in Table~\ref{tab:experimental-parameters}.

Figure~\ref{fig:standard-rc-results}(a) shows the eigenvalue distribution of the effective reservoir transition matrix $\mathbf{A}_{\mathrm{eff}}$ (in this preliminary experiment, $\alpha = 1$, and thus $\mathbf{A}_{\mathrm{eff}} = \mathbf{A}$). Because the observation dimension is $D=1$, the effective closed-loop matrix in the prediction phase, $\mathbf{W}_{\mathrm{cl,eff}} = \mathbf{A}_{\mathrm{eff}} + \alpha \mathbf{W}_{\mathrm{in}}\mathbf{W}_{\mathrm{out}}$, can be regarded as $\mathbf{A}_{\mathrm{eff}}$ subjected to the rank-one perturbation $\alpha \mathbf{W}_{\mathrm{in}}\mathbf{W}_{\mathrm{out}}$. Figure~\ref{fig:standard-rc-results}(b) shows a typical eigenvalue distribution of $\mathbf{W}_{\mathrm{cl,eff}}$. As a result of the perturbation, several slow eigenvalues emerge outside the circle of radius $\rho(\mathbf{A}_{\mathrm{eff}}) = 0.5$.

Figure~\ref{fig:standard-rc-results}(c) shows the reservoir state sequence $\mathbf{R}_T$ obtained during the listening phase, projected onto its first three principal components. The two-lobe structure characteristic of the Lorenz attractor is geometrically reconstructed in the reservoir state space. Figure~\ref{fig:standard-rc-results}(d) shows the result of the prediction phase (closed-loop prediction). As a measure of prediction performance, we adopt the valid prediction time (VPT)~\cite{Vlachas2020-ob}, the normalized elapsed time (measured in Lyapunov times) until the normalized root-mean-square error (NRMSE) first exceeds a threshold (here 0.5). Letting $\hat{\bm{u}}_t$ denote the prediction at time $t$, the NRMSE at time $t$ and the VPT are defined respectively as
\begin{equation}
    \mathrm{NRMSE}(\hat{\bm{u}}_t) = \sqrt{\frac{1}{D}
    \sum_{i=1}^{D} \frac{(\hat{u}_{t,i} - u_{t,i})^2}{\sigma_i^2}},
    \label{eq:nrmse}
\end{equation}
\begin{equation}
    \mathrm{VPT} = \lambda_{\max}^{-1}
    \max \{ t_f : \mathrm{NRMSE}(\hat{\bm{u}}_t) < 0.5, \
    \forall\, t \le t_f \}, \label{eq:vpt}
\end{equation}
where $\sigma_i^2$ is the variance of the $i$-th component of the observation sequence and $\lambda_{\max}$ is the largest Lyapunov exponent of the target system. The vertical lines in Fig.~\ref{fig:standard-rc-results}(d) mark the valid prediction times. The mean VPT over 20 independent realizations of the RC is 5.11 Lyapunov times, but it varies widely across realizations, from 0.87 to 8.69 Lyapunov times.

\begin{figure*}[htbp]
  \centering
  \includegraphics[width=\textwidth]{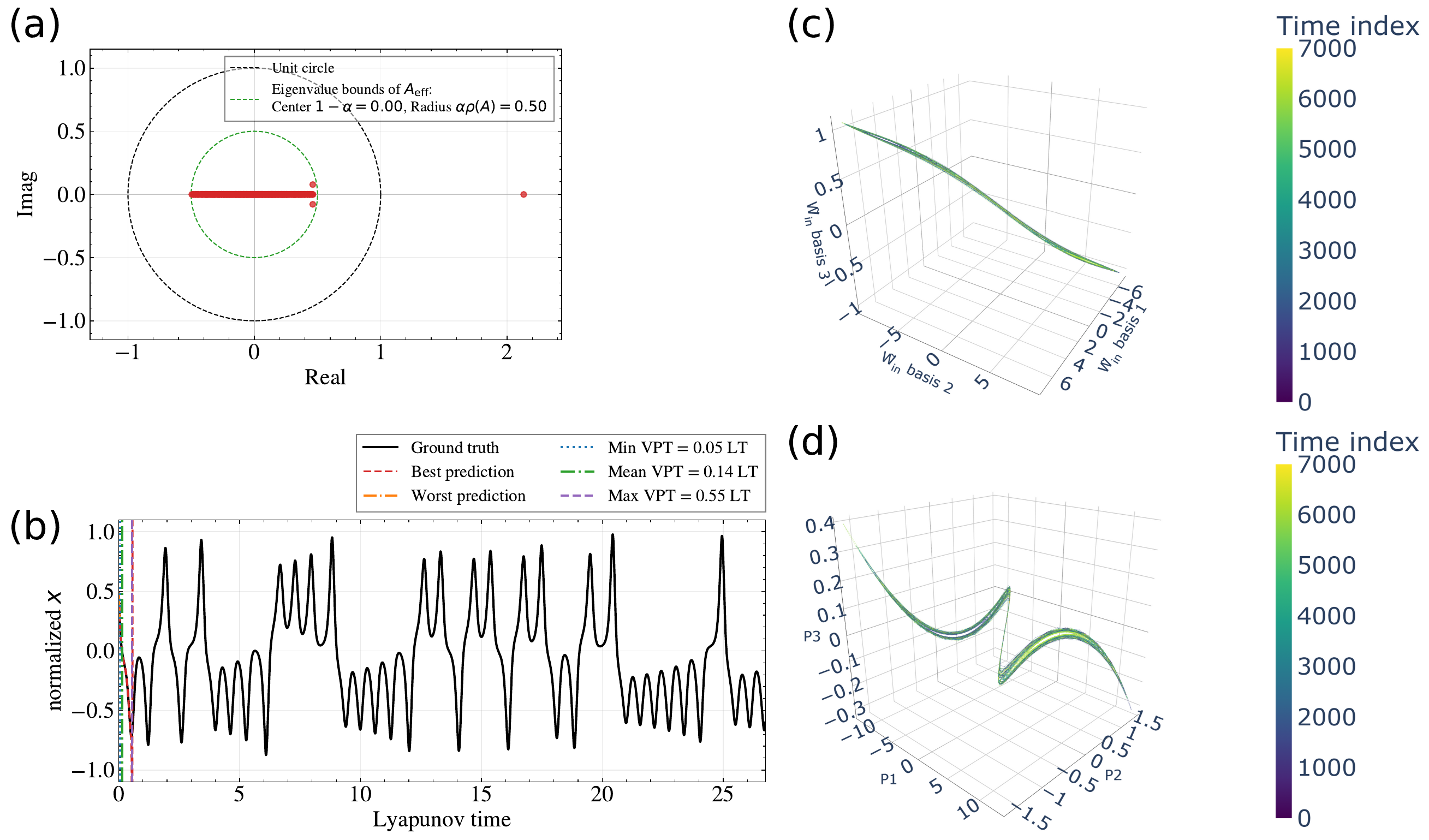}
\caption{Deterministic input-layer RC without slow-spectrum preshaping for the Lorenz-63 system. (a) Eigenvalue distribution of $\mathbf{W}_{\mathrm{cl,eff}}$. Of the three eigenvalues moved from those of $\mathbf{A}_{\mathrm{eff}}$, two lie inside the circle of radius $\rho(\mathbf{A}_{\mathrm{eff}}) = 0.5$, and the third escapes along the real axis to outside the unit circle. (b) Failure of closed-loop prediction. (c) Listening-phase reservoir states projected onto the designed subspace $\operatorname{col}(\tilde{\mathbf{W}}_{\mathrm{in}})$, the column space of $\tilde{\mathbf{W}}_{\mathrm{in}}$, where the attractor collapses to a filament from any viewpoint. (d) The same states projected onto a different three-dimensional PCA subspace, in which the two-lobe topology is recovered.}
  \label{fig:deterministic-win-results}
\end{figure*}

A critical issue arises in the long-term statistical behavior of the closed-loop RC, as revealed by its Lyapunov spectrum [Fig.~\ref{fig:standard-rc-results}(e)]. The Lorenz attractor possesses one positive, one zero, and one negative Lyapunov exponent, so if the RC successfully reproduced the target dynamics, the first three Lyapunov exponents computed from Eq.~\eqref{eq:closed-loop-dynamics} should match these values. In practice, the second Lyapunov exponent of the closed-loop RC, which should be zero, falls  within $[-0.1, 0.1]$ in only 3 of 20 trials and tends to take excessively large positive values. This strongly implies that one of the transverse Lyapunov exponents is positive, meaning that the reconstructed invariant set in the reservoir state space is transversally unstable under the closed-loop dynamics. In Ref.~\onlinecite{Oishi2026-fz}, such transverse instability was attributed to spurious slow eigenvalues of the effective closed-loop matrix $\mathbf{W}_{\mathrm{cl,eff}}$. The emergence of several slow eigenvalues observed in Fig.~\ref{fig:standard-rc-results}(b) is consistent with this mechanism.

The preliminary results of Fig.~\ref{fig:standard-rc-results} show that dynamical-system reproduction by standard RC depends strongly on the random initialization of the internal weights and can easily fail to generate long-term chaotic trajectories due to transverse instability. Because the closed-loop prediction can fail abruptly even if hyperparameters are tuned using various optimization methods~\cite{Yperman2016-xd, Bala2018-zq, Racca2021-xx, Platt2023-lh}, this poor reproducibility across initializations has been the main obstacle to establishing standard RC as a reliable tool for time-series analysis and chaotic prediction.

\subsection{Deterministic input-layer design for ARAS and its failure under partial observation}
\label{sec:deterministic-win-limitations}

To overcome the transverse instability caused by random initialization in standard RC, we recently proposed a task-agnostic approach that designs the input layer $\mathbf{W}_{\mathrm{in}}$ deterministically to realize Attractor Reconstruction in Attracting Submanifolds (ARAS)~\cite{Oishi2026-fz}. ARAS requires the reconstructed invariant set to lie in a transversally attracting submanifold of the representation space. To implement ARAS via input-layer design, one constructs $\mathbf{W}_{\mathrm{in}}$ such that its columns lie in a specified $D'$-dimensional real $\mathbf{A}_{\mathrm{eff}}$-invariant subspace. Specifically, let $\tilde{\mathbf{W}}_{\mathrm{in}} \in \mathbb{R}^{N \times D'}$ be a matrix whose columns form a basis of this subspace. Since the subspace is $\mathbf{A}_{\mathrm{eff}}$-invariant, there exists a coefficient matrix $\mathbf{X} \in \mathbb{R}^{D' \times D'}$ such that $\mathbf{A}_{\mathrm{eff}} \tilde{\mathbf{W}}_{\mathrm{in}} = \tilde{\mathbf{W}}_{\mathrm{in}} \mathbf{X}$. Since every column of $\mathbf{W}_{\mathrm{in}}$ belongs to this subspace, we can write $\mathbf{W}_{\mathrm{in}} = \tilde{\mathbf{W}}_{\mathrm{in}} \mathbf{B}$ for some coefficient matrix $\mathbf{B} \in \mathbb{R}^{D' \times D}$. Then $\tilde{\mathbf{W}}_{\mathrm{in}}$ also spans an invariant subspace of the effective closed-loop matrix $\mathbf{W}_{\mathrm{cl,eff}} = \mathbf{A}_{\mathrm{eff}} + \alpha \mathbf{W}_{\mathrm{in}}\mathbf{W}_{\mathrm{out}}$:
\begin{align}
\mathbf{W}_{\mathrm{cl,eff}} \tilde{\mathbf{W}}_{\mathrm{in}}
&= (\mathbf{A}_{\mathrm{eff}}
+ \alpha \tilde{\mathbf{W}}_{\mathrm{in}}
\mathbf{B}\mathbf{W}_{\mathrm{out}})
\tilde{\mathbf{W}}_{\mathrm{in}} \nonumber \\
&= \tilde{\mathbf{W}}_{\mathrm{in}}
(\mathbf{X} + \alpha \mathbf{B} \mathbf{W}_{\mathrm{out}}
\tilde{\mathbf{W}}_{\mathrm{in}}).
\label{eq:closed-loop-invariance}
\end{align}
With this design, the number of eigenvalues of $\mathbf{A}_{\mathrm{eff}}$ that can be moved by the perturbation $\alpha \mathbf{W}_{\mathrm{in}}\mathbf{W}_{\mathrm{out}}$ to become slow modes is strictly controlled: it is up to $D'$ (see also  Theorem~1 of Ref.~\onlinecite{Oishi2026-fz}). Importantly, $\tilde{\mathbf{W}}_{\mathrm{in}}$ is the invariant subspace associated with these potentially slow eigenvalues of $\mathbf{W}_{\mathrm{cl,eff}}$. Intuitively, keeping this subspace invariant with respect to both $\mathbf{A}_{\mathrm{eff}}$ and $\mathbf{W}_{\mathrm{cl,eff}}$ helps ensure that the reconstructed invariant set formed during the listening phase remains invariant under the closed-loop dynamics [Fig.~\ref{fig:concept}(b)]. Furthermore, the advantage of this design goes beyond this geometric consistency. Crucially, all remaining eigenvalues stay at those of $\mathbf{A}_{\mathrm{eff}}$ and can be guaranteed to be strongly contracting. Due to this spectral separation between the potentially slow modes and the strongly contracting transverse modes, Fenichel theory~\cite{Kuehn2015-pl, Fenichel1979-ej, Hirsch1977-ry} suggests the existence of a locally attracting slow invariant manifold $\mathscr{M}_{\mathrm{slow}}$ whose dimension is at most $D'$. By appropriately setting $D'$ close to the dimension of the target invariant set, one restricts the maximum dimension of this resulting slow manifold in the reservoir state space. Thus, by suppressing the emergence of spurious slow eigenvalues in $\mathbf{W}_{\mathrm{cl,eff}}$, this approach reduces the risk of transversally unstable modes emerging from the excess dimensions of the slow manifold. Our previous study showed that this deterministic input-layer design removes the influence of the random initialization of $\mathbf{A}$ almost entirely and reproduces dynamical systems accurately over a wide range of hyperparameters under full observation.

This success, however, is limited to the full-observation setting. In the partial-observation setting, we found that this input-layer design alone fails: the submanifold designed to be attracting cannot retain enough observation history, and the reconstructed invariant set collapses to an excessively low-dimensional structure.

We repeated the Lorenz-63 partial-observation experiment of Sec.~\ref{sec:standard-rc-limitations} ($D = 1$) with $\mathbf{W}_{\mathrm{in}}$ constructed from a three-dimensional real $\mathbf{A}_{\mathrm{eff}}$-invariant subspace. The results are shown in Fig.~\ref{fig:deterministic-win-results}. Figure~\ref{fig:deterministic-win-results}(a) shows the eigenvalue distribution of $\mathbf{W}_{\mathrm{cl,eff}}$ obtained with the same $\mathbf{A}$ as in Fig.~\ref{fig:standard-rc-results}(a). Because $\mathbf{W}_{\mathrm{in}}$ is built from the invariant subspace, up to three eigenvalues can move from those of $\mathbf{A}_{\mathrm{eff}}$. Under partial observation, however, we observe that only one of these eigenvalues is substantially perturbed away from the fast-contracting spectrum, and it escapes to a large positive value on the real axis. In the successful full-observation cases, the three moved eigenvalues were placed near the unit circle (see Ref.~\onlinecite{Oishi2026-fz}). Thus, under partial observation, an insufficient number of modes are released from fast contraction to form the required slow submanifold, and the released modes can instead become strongly unstable, lying far outside the unit circle. Consequently, closed-loop prediction [Fig.~\ref{fig:deterministic-win-results}(b)] fails immediately and does not reproduce the chaotic trajectory at all. When the actual listening-phase reservoir states are projected onto $\operatorname{col}(\tilde{\mathbf{W}}_{\mathrm{in}})$, the subspace of the reservoir state space spanned by the columns of $\tilde{\mathbf{W}}_{\mathrm{in}}$, the two-lobe topology of the attractor completely collapses into a one-dimensional, filament-like structure [Fig.~\ref{fig:deterministic-win-results}(c)]. This low dimensionality reflects the fact that only one eigenvalue is no longer fast. By contrast, projecting the reservoir states onto a different, suitably chosen subspace [Fig.~\ref{fig:deterministic-win-results}(d)] reveals a thin, barely recognizable two-lobe topology. The attractor is indeed reconstructed in the reservoir state space, as expected from the ESN embedding conjecture~\cite{Hart2020-kx}. However, what is designed to be attracting is the slow invariant manifold $\mathscr{M}_{\mathrm{slow}}$ approximated by $\operatorname{col}(\tilde{\mathbf{W}}_{\mathrm{in}})$, which retains only one effective dimension and is thus insufficient for reconstructing the target attractor.

The reason the attractor topology cannot be reconstructed on the intended slow invariant manifold $\mathscr{M}_{\mathrm{slow}}$ can be understood as follows. To induce the fast--slow separation, the spectral radius of $\mathbf{A}_{\mathrm{eff}}$ is set sufficiently small. Therefore, in the open-loop update of Eq.~\eqref{eq:rc-update}, the current input $\mathbf{W}_{\mathrm{in}}\bm{u}_t$ dominates the rapidly decaying recurrent term $\mathbf{A}\bm{r}_t$, which carries the input history. As a result, the observation history is not retained sufficiently to unfold the data into the intended $D'$-dimensional slow structure. Instead, the reservoir states collapse onto the $D$-dimensional structure associated with the columns of $\mathbf{W}_{\mathrm{in}}$ [Fig.~\ref{fig:deterministic-win-results}(c)], which constitutes a collapsed version of the slow invariant manifold $\mathscr{M}_{\mathrm{slow}}$. Even if the reservoir states during the listening phase form an embedding of the target attractor, the resulting embedded set cannot be contained in the collapsed $\mathscr{M}_{\mathrm{slow}}$. Consequently, once the reservoir is operated in closed loop, trajectories can leave the reconstructed invariant set, so the target dynamics cannot be reproduced during the prediction phase.

\section{Proposed method: slow-spectrum preshaping}
\label{sec:proposed-methodology}
Our goal is to establish an approach that achieves ARAS deterministically and task-agnostically under partial observation, similar to what the input-layer design of Ref.~\onlinecite{Oishi2026-fz} accomplished under full observation. However, as discussed in the previous section, directly applying this ARAS-oriented input-layer design under partial observation is insufficient because the reconstruction within the designed submanifold collapses into an extremely low-dimensional structure [Fig.~\ref{fig:concept}(c)]. Interestingly, we found that lifting the input dimension $D$ by using delay coordinates of the observation still failed to achieve ARAS (see Appendix~\ref{sec:delay-coordinates}). This suggests that overcoming the challenge of partial observation requires shaping the intrinsic reservoir dynamics itself, rather than merely adjusting the input layer.

\begin{figure*}[t]
  \centering
  \includegraphics[width=\textwidth]{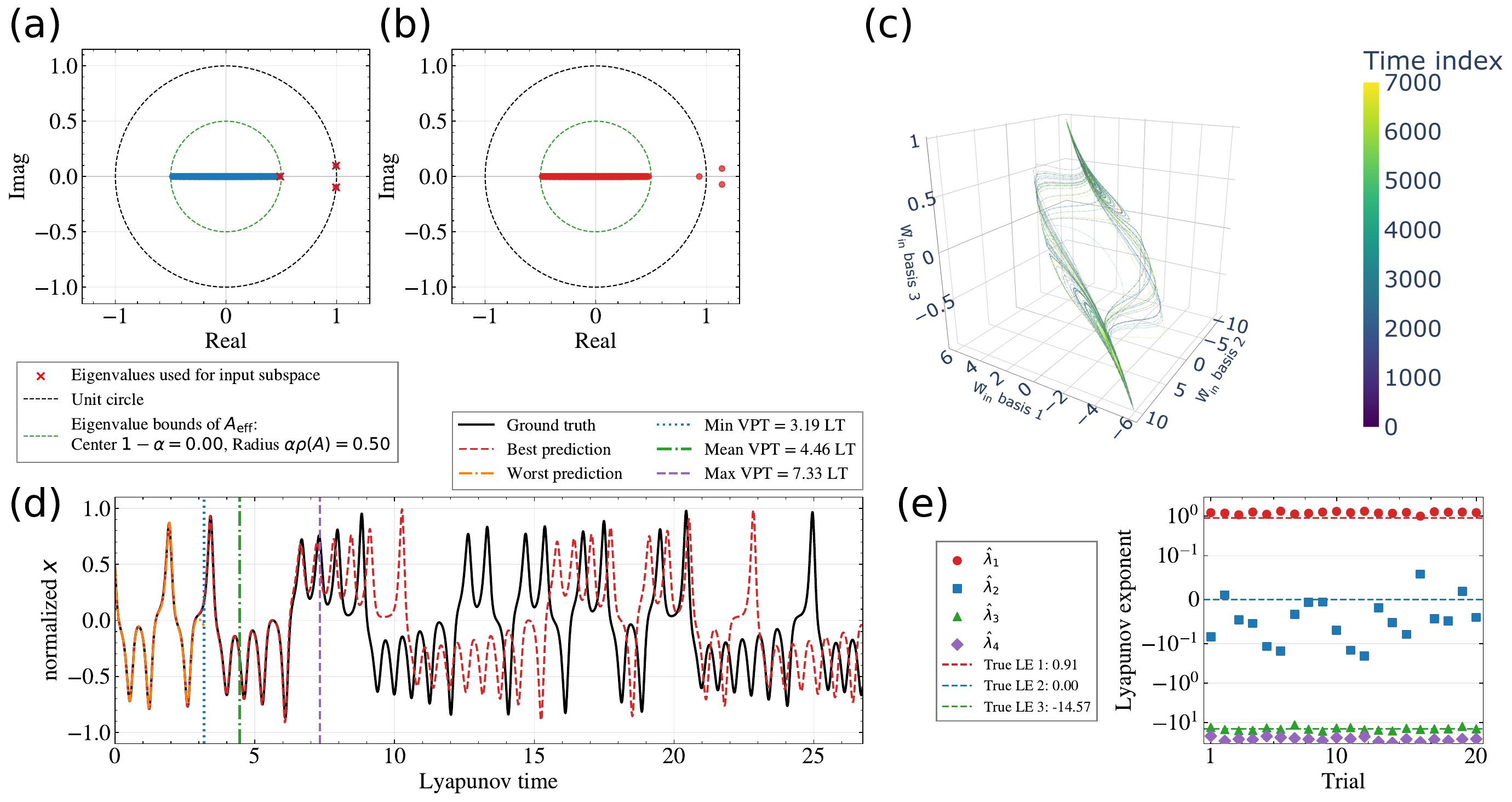}
\caption{Deterministic input-layer RC with slow-spectrum preshaping for the Lorenz-63 system. (a) Eigenvalue distribution of $\mathbf{A}'_{\mathrm{eff}}$ after slow-spectrum preshaping. (b) Eigenvalue distribution of $\mathbf{W}_{\mathrm{cl,eff}}$. (c) Listening-phase reservoir states projected onto the designed subspace $\operatorname{col}(\tilde{\mathbf{W}}_{\mathrm{in}})$, recovering the two-lobe topology. (d) Short-term prediction performance. (e) Lyapunov spectrum of the closed-loop RC, plotted on a symmetric-log scale that is linear within $[-0.1, 0.1]$; the ground-truth values of the Lorenz attractor are indicated by horizontal dashed lines.}
  \label{fig:slow-spectrum-preshaping-results}
\end{figure*}

To realize ARAS under partial observation, we introduce \textit{slow-spectrum preshaping}: before constructing the input layer, we place a small number of slow eigenvalues in the reservoir transition matrix (see also Fig.~\ref{fig:slow-spectrum-preshaping-results}(a)), and then build $\mathbf{W}_{\mathrm{in}}$ from a $D'$-dimensional invariant subspace associated with these slow eigenvalues. The design rationale is twofold. First, the slow invariant manifold $\mathscr{M}_{\mathrm{slow}}$ approximated by $\operatorname{col}(\tilde{\mathbf{W}}_{\mathrm{in}})$ retains a memory of past inputs during the listening phase governed by Eq.~\eqref{eq:rc-update}. It  thereby unfolds the observation history within $\mathscr{M}_{\mathrm{slow}}$, reconstructing the topology in the same way as a delay-coordinate embedding. Second, because $\mathbf{W}_{\mathrm{in}}$ is constructed from $\tilde{\mathbf{W}}_{\mathrm{in}}$, this submanifold can be kept transversally attracting under the closed-loop dynamics, so the attractor reconstruction is guided into an attracting slow submanifold, as discussed in Sec.~\ref{sec:deterministic-win-limitations}.

A reader familiar with RC may be concerned that placing slow eigenvalues in the reservoir transition matrix destroys the contraction property of $f$, i.e., that the ESP no longer holds. However, learning a chaotic system with RC requires only that the reservoir state lose its dependence on the initial state for the specific input sequence generated by the target system, not for the arbitrary input sequences that the standard ESP demands~\cite{Manjunath2013-bj}. We thus address this concern in Sec.~\ref{sec:esp-verification} by computing the CLEs of Eq.~\eqref{eq:rc-update} and confirming that GS is maintained for reasonable placements of the slow eigenvalues.

\subsection{Slow-spectrum preshaping procedure}
\label{sec:slow-spectrum-preshaping-details}

We first describe the procedure in detail. Algorithm~\ref{alg:slow-spectrum-preshaping} constructs the preshaped reservoir transition matrix $\mathbf{A}'$ by assigning the specified slow eigenvalues while preserving the eigenstructure of the initially generated matrix $\mathbf{A}$ as much as possible.\footnote{Preserving the eigenstructure is not essential to the proposed framework. This conservative construction is adopted here to isolate the effect of slow-spectrum preshaping. In practical implementations, simpler constructions would suffice for slow-spectrum preshaping while allowing the eigenvectors to change.} In Step~2 of Algorithm~\ref{alg:slow-spectrum-preshaping}, which eigenvalues of $\mathbf{A}$ to replace may be chosen arbitrarily. In our experiments (not shown), the choice of eigenvalues did not substantially affect the performance of the proposed method.

\begin{algorithm}[p]
\caption{Slow-spectrum preshaping of the reservoir transition matrix}
\label{alg:slow-spectrum-preshaping}
\begin{algorithmic}[1]
\Require $\mathbf{A} \in \mathbb{R}^{N \times N}$, specified eigenvalues
$\hat{\mathcal{M}} = (\hat{\mu}_1, \dots, \hat{\mu}_k)$ (ordered
sequence where complex eigenvalues appear in adjacent conjugate pairs:
$\hat{\mu}_{i+1} = \bar{\hat{\mu}}_i$).
\Ensure Preshaped reservoir transition matrix $\mathbf{A}' \in
\mathbb{R}^{N \times N}$.
\State Compute eigenvalues $\mu_1, \dots, \mu_N$ of $\mathbf{A}$,
right eigenvector matrix $\mathbf{V} = [\bm{v}_1, \dots, \bm{v}_N]$ with unit-norm columns,
and left eigenvector matrix $\mathbf{U} = [\bm{u}_1, \dots, \bm{u}_N]^{\top} = \mathbf{V}^{-1}$
\State $(\mu_1, \dots, \mu_k) \gets$ select $k$ eigenvalues of
$\mathbf{A}$ to be replaced
\Statex \textit{Note: Complex eigenvalues must be selected in conjugate
pairs.}
\State $\mathbf{X}, \mathbf{Y} \in \mathbb{R}^{N \times k} \gets$ real
bases of the right and left invariant subspaces of $\mathbf{A}$
associated with the selected eigenvalues, with $\mathbf{Y}^{\top}
\mathbf{X} = \mathbf{I}_k$
\Statex \textit{Note: We form $\mathbf{X}$ from the selected right
eigenvectors, using the real and imaginary parts for a conjugate pair.}
\State $\hat{\mathbf{B}} \in \mathbb{R}^{k \times k} \gets$ real
canonical form associated with the specified eigenvalues
$\hat{\mathcal{M}}$
\Statex \textit{Note: $\hat{\mathbf{B}}$ is block diagonal with the
block $[\hat{\mu}_i]$ for a real specified eigenvalue and
$\begin{bmatrix} a & b \\ -b & a \end{bmatrix}$ for a specified
conjugate pair $a \pm \mathrm{i}b$.}
\State $\mathbf{A}' \gets \mathbf{A} + \mathbf{X} (\hat{\mathbf{B}} -
\mathbf{Y}^{\top} \mathbf{A} \mathbf{X}) \mathbf{Y}^{\top}$
\State \Return $\mathbf{A}'$
\end{algorithmic}
\end{algorithm}

Algorithm~\ref{alg:deterministic-win} describes the deterministic input-layer design of Ref.~\onlinecite{Oishi2026-fz}, adjusted to the leaky-integrator setting. As noted in Sec.~\ref{sec:rc-fundamentals}, the RC update rule includes a leaky integrator, so the linearized dynamics near the origin are governed by the effective reservoir transition matrix $\mathbf{A}'_{\mathrm{eff}} = (1-\alpha)\mathbf{I} + \alpha \mathbf{A}'$ (when the bias $\bm{b}$ is zero or sufficiently small). The proposed method therefore constructs $\mathbf{W}_{\mathrm{in}}$ from a real invariant subspace of this $\mathbf{A}'_{\mathrm{eff}}$.\footnote{The slow eigenvalues are placed in $\mathbf{A}'$ by Algorithm~\ref{alg:slow-spectrum-preshaping}, not directly in $\mathbf{A}'_{\mathrm{eff}}$. Because $\mathbf{A}'_{\mathrm{eff}} = (1-\alpha)\mathbf{I} + \alpha\mathbf{A}'$ shares its eigenvectors with $\mathbf{A}'$, an eigenvalue $\mu$ of $\mathbf{A}'$ corresponds to the eigenvalue $1-\alpha+\alpha\mu$ of $\mathbf{A}'_{\mathrm{eff}}$. The leaking rate $\alpha$ therefore shifts the placed eigenvalues, but the associated invariant subspace is unchanged and the eigenvalues remain slow, so the goal of introducing a few slow eigenvalues into $\mathbf{A}'_{\mathrm{eff}}$ is still achieved.} Particularly important is that $\mathbf{W}_{\mathrm{in}}$ is built from $\tilde{\mathbf{W}}_{\mathrm{in}}$, the $D'$-dimensional invariant subspace associated with the slow eigenvalues placed by Algorithm~\ref{alg:slow-spectrum-preshaping}. In this way, the slow spectral submanifold approximated by the subspace $\operatorname{col}(\tilde{\mathbf{W}}_{\mathrm{in}})$ is kept transversally attracting under the closed-loop dynamics. The scaling in step~\ref{step:scaling} of Algorithm~\ref{alg:deterministic-win} assumes that the input signal is normalized to the range $[-1, 1]$. \footnote{When such normalization is not applicable, an appropriate scaling factor should be selected individually according to the amplitude and range of the target signal.} With $\mathbf{A}'$ and $\mathbf{W}_{\mathrm{in}}$ constructed by the above procedures, the output layer $\mathbf{W}_{\mathrm{out}}$ is optimized following the same procedure as for standard RC described in Sec.~\ref{sec:rc-fundamentals}.

\begin{algorithm}[htbp]
\caption{Deterministic input-layer initialization}
\label{alg:deterministic-win}
\begin{algorithmic}[1]
\Require $\mathbf{A}' \in \mathbb{R}^{N \times N}$ (from
Algorithm~\ref{alg:slow-spectrum-preshaping}), input dimension $D$,
invariant subspace dimension $D' (\geq D)$, leaking rate $\alpha$,
input scaling ratio $\gamma$.
\Ensure Deterministic input-layer matrix $\mathbf{W}_{\mathrm{in}}
\in \mathbb{R}^{N \times D}$.
\State $\mathbf{A}'_{\mathrm{eff}} \gets (1-\alpha)\mathbf{I} + \alpha
\mathbf{A}'$ \Comment{Linearized transition matrix}
\State Compute eigenvalues $\mu^{\mathrm{eff}}_1, \dots,
\mu^{\mathrm{eff}}_N$ of $\mathbf{A}'_{\mathrm{eff}}$ and their right
eigenvectors $\mathcal{V}_{\mathrm{eff}}$
\State Let $\mathcal{M}_{D'}$ be the set of $D'$ eigenvalues with the
largest absolute values, where complex eigenvalues must be taken in conjugate
pairs
\Statex \textit{Note: Always include the preshaped slow-eigenvalue
subset $\hat{\mathcal{M}}$.}
\State $\mathcal{V}_{D'} \gets$ eigenvectors in
$\mathcal{V}_{\mathrm{eff}}$ corresponding to $\mathcal{M}_{D'}$
\State Initialize an empty basis matrix $\mathbf{W} \in \mathbb{R}^{N
\times D'}$ with $N$ rows
\For{each $\bm{v} \in \mathcal{V}_{D'}$}
    \If{$\bm{v} \in \mathbb{R}^N$}
        \State Append $\bm{v}$ as a column to $\mathbf{W}$
    \ElsIf{$\bm{v}, \bar{\bm{v}}$ is a conjugate pair}
        \State Append $\mathrm{Re}(\bm{v})$ and $\mathrm{Im}(\bm{v})$
        as columns to $\mathbf{W}$
    \EndIf
\EndFor
\State $\mathbf{Q}, \mathbf{R} \gets \mathrm{qr}(\mathbf{W})$
\Comment{Orthonormalize the basis}
\State Sample coefficient matrix $\mathbf{C} \in \mathbb{R}^{D' \times
D}$ with $C_{ij} \overset{\text{i.i.d.}}{\sim} \mathcal{N}(0, 1)$
\State $\mathbf{W}_{\mathrm{in}} \gets \mathbf{Q} \mathbf{C}$
\For{$i = 1 \dots D$} \Comment{Normalize each column $\bm{w}_i$ of
$\mathbf{W}_{\mathrm{in}}$}
    \State $\bm{w}_i \gets \bm{w}_i / \|\bm{w}_i\|_2$
\EndFor
\State \label{step:scaling} $\mathbf{W}_{\mathrm{in}} \gets \gamma
\sqrt{N/D}\, \mathbf{W}_{\mathrm{in}}$ \Comment{Input scaling}
\State \Return $\mathbf{W}_{\mathrm{in}}$
\end{algorithmic}
\end{algorithm}

In the proposed method, the number $k$ of slow eigenvalues to introduce is an important parameter. It can be regarded as the number of dimensions reserved for retaining past history. Setting the dimension of the invariant subspace $\tilde{\mathbf{W}}_{\mathrm{in}}$ to $D' = k + 1$ is analogous to constructing a $D'$-dimensional delay-coordinate vector $(u_t, u_{t-\tau_1}, \dots, u_{t-\tau_k})$ from the current observation and $k$ delayed observations. How to select an appropriate embedding dimension $D'$ has been discussed extensively in delay-embedding theory. According to the embedding theorems~\cite{Takens1981-mw, Sauer1991-ew} (as in the weak Whitney embedding theorem~\cite{Whitney1936-es}), a manifold of dimension $d$ can generically be embedded by taking $D' \geq 2d + 1$. Conversely, choosing an excess dimension causes problems known as overembedding \cite{Kantz1997-lk,Olbrich1997-nx,Sauer1998-la,Pecora2007-fj,Pecora2025-js}, including failures in Lyapunov-spectrum estimation, overfitting, and degraded time-series analysis.

It is therefore reasonable to identify the minimal embedding dimension $D'$ required for the target system and to set $k = D' - 1$. Determining $D'$ \textit{a priori} for an unknown system is not trivial, but data-driven estimates are available, such as the false-nearest-neighbor method \cite{Kennel1992-rz} and mutual-information-based criteria \cite{Liebert1991-dz}. Since the experiments in this paper target the Lorenz system, we set $D' = 3$ (i.e., $k = 2$), which we found to be the minimal sufficient choice.

\subsection{Demonstration on the partially observed Lorenz-63 system}
\label{sec:slow-spectrum-preshaping-results}

To examine how the deterministic input-layer design combined with slow-spectrum preshaping reconstructs the attractor under partial observation, we repeat the experiment of Sec.~\ref{sec:standard-rc-limitations} (only the $x$ variable of the Lorenz-63 system is observed) and compare the results.

Starting from the randomly generated reservoir transition matrix $\mathbf{A}$ whose eigenvalue distribution is shown in Fig.~\ref{fig:standard-rc-results}(a), we applied slow-spectrum preshaping that moved two eigenvalues to $\cos(\pi/32) \pm \mathrm{i}\sin(\pi/32)$ while preserving the original spectrum and invariant subspaces as much as possible [Fig.~\ref{fig:slow-spectrum-preshaping-results}(a), Algorithm~\ref{alg:slow-spectrum-preshaping}]. We then set $D' = 3$, matching the dimension of the Lorenz system, and constructed the deterministic input layer $\mathbf{W}_{\mathrm{in}}$ from the real invariant subspace of the resulting effective transition matrix $\mathbf{A}'_{\mathrm{eff}} = (1-\alpha)\mathbf{I} + \alpha \mathbf{A}'$ using Algorithm~\ref{alg:deterministic-win}.

With the $\mathbf{A}'$ and $\mathbf{W}_{\mathrm{in}}$ thus obtained, we trained the output layer to predict the $x$ variable of the Lorenz system and built the closed-loop system. Figure~\ref{fig:slow-spectrum-preshaping-results}(b) shows the eigenvalue distribution of the trained closed-loop matrix $\mathbf{W}_{\mathrm{cl,eff}}$. Only the three eigenvalues of $\mathbf{A}'_{\mathrm{eff}}$ selected to construct $\tilde{\mathbf{W}}_{\mathrm{in}}$ are perturbed and move to the vicinity of the unit circle, while all remaining eigenvalues are unaffected by training.

Figure~\ref{fig:slow-spectrum-preshaping-results}(c) shows the listening-phase reservoir states projected onto the slow spectral subspace $\operatorname{col}(\tilde{\mathbf{W}}_{\mathrm{in}})$. Without preshaping, this projection collapses to a one-dimensional filament [Fig.~\ref{fig:deterministic-win-results}(c)]. With the placed slow eigenvalues, by contrast, the observation history is retained and the Lorenz two-lobe topology is clearly reconstructed [Fig.~\ref{fig:slow-spectrum-preshaping-results}(c)].

\begin{table*}[htbp]
\centering
\caption{Experimental parameters used in the numerical simulations. $\mathbf{1}$ denotes the $N$-dimensional all-ones vector.}
\label{tab:experimental-parameters}
\begin{tabular}{llccc}
\hline\hline
RC hyperparameters & Symbol & Value (Figs.~\ref{fig:standard-rc-results}--\ref{fig:slow-spectrum-preshaping-results}) & Value
(Secs.~\ref{sec:esp-verification},
\ref{sec:attractor-reconstruction-performance}) & Value
(Sec.~\ref{sec:parameter-robustness-evaluation}) \\
\hline
Reservoir size & $N$ & 300 & 300 & 300 \\
Spectral radius & $\rho(\mathbf{A})$ & 0.5 & $[0.1, 0.3, 0.5, 0.7,
0.9]$ & $[0.1, 0.3, 0.5, 0.7, 0.9]$ \\
Leaking rate & $\alpha$ & 1.0 & 1.0 & $[0.2, 0.4, 0.6, 0.8, 1.0]$ \\
Input bias & $\bm{b}$ & $0.01 \times \mathbf{1}$ & $0.01 \times
\mathbf{1}$ & $0.01 \times \mathbf{1}$ \\
Input scaling ratio & $\gamma$ & $1.0$ & $1.0$ & $[0.5, 0.75, 1.0,
1.25, 1.5]$ \\
Regularization & $\beta$ & 0 & 0 & $[0.0, 10^{-24}, 10^{-20},
10^{-16}, 10^{-12}]$ \\
\hline
Proposed method parameters &  &  & &  \\
Number of slow eigenvalues & $k$ & 2 & 2 & 2 \\
Dimension of $\tilde{\mathbf{W}}_{\mathrm{in}}$ & $D'$ & 3 & 3 & 3 \\
\hline
Experimental settings &  &  & &  \\
Warm-up steps & $T_\mathrm{w}$ & 1000 & 1000 & 1000 \\
Training steps & $T$ & 2000 & 2000 & 2000 \\
Test points & -- & -- & -- & 50 \\
\hline
\end{tabular}
\end{table*}

Figure~\ref{fig:slow-spectrum-preshaping-results}(d) shows the prediction phase of this closed-loop system. The mean valid prediction time over 20 independent realizations of the RC is 4.46 Lyapunov times, slightly shorter than the 5.11 Lyapunov times of the standard RC in Fig.~\ref{fig:standard-rc-results}(d). However, the shortest valid prediction time among the realizations is 3.19 Lyapunov times, much longer than the 0.87 Lyapunov times of the standard RC. Furthermore, in the estimated Lyapunov spectra shown in Fig.~\ref{fig:slow-spectrum-preshaping-results}(e), the second Lyapunov exponent falls within $[-0.1, 0.1]$ in 16 of 20 trials and never becomes large and positive, avoiding the transverse instability ($\mathrm{TLE} > 0$) observed for standard RC. These results support the claim that slow-spectrum preshaping keeps the reconstructed invariant set within the attracting submanifold even under partial observation and thereby enables stable long-term emulation of chaotic dynamics.

\section{Comprehensive evaluation}
\label{sec:comprehensive-evaluation}

This section systematically examines the effectiveness and robustness of the proposed method over a broader range of hyperparameter settings. The initial experiment of Sec.~\ref{sec:slow-spectrum-preshaping-results} showed promising results, but establishing the generality of the method requires answering two questions. The first is whether deliberately introducing slow eigenvalues into the reservoir transition matrix $\mathbf{A}'$ violates the echo state property. This is addressed in Sec.~\ref{sec:esp-verification}, where we compute CLEs and verify quantitatively that generalized synchronization is maintained during the listening phase. The second is how the performance depends on the placement of the preshaped eigenvalues and on the RC hyperparameters. The placement $\cos(\pi/32) \pm \mathrm{i}\sin(\pi/32)$ in Sec.~\ref{sec:slow-spectrum-preshaping-results} was an arbitrary choice, and the dependence on it must be assessed. To this end, Secs.~\ref{sec:esp-verification}, \ref{sec:attractor-reconstruction-performance}, and \ref{sec:parameter-robustness-evaluation} repeat the experiments with the six eigenvalue settings
\begin{align}
    \text{Cases 1--3:} & \;\hat{\mu}_{1}, \hat{\mu}_{2} = \cos\theta \pm
    \mathrm{i}\sin\theta,
    \; \theta \in \{ \pi/16, \pi/32, \pi/64 \}, \nonumber \\
    \text{Cases 4--6:} & \;\hat{\mu}_1 \in \{ 1.01, 1.05, 1.1 \},
    \; \hat{\mu}_2 = 1/\hat{\mu}_1,
    \label{eq:target-eigenvalue-settings}
\end{align}
and show that, compared with standard RC, the proposed method improves the prediction performance and the capability to reproduce chaos over a much wider parameter region. The two groups differ in the type of slow spectrum they introduce: Cases 1--3 place a complex-conjugate pair on the unit circle (a rotation-like placement, as used in Sec.~\ref{sec:slow-spectrum-preshaping-results}), whereas Cases 4--6 use a reciprocal real pair $\hat{\mu}_2 = 1/\hat{\mu}_1$ (a saddle-like placement).

The common experimental settings for these experiments are summarized in Table~\ref{tab:experimental-parameters}. In all experiments of this paper, the observation sequence $\bm{u}_t$ from the target system is normalized in advance to the range $[-1, 1]$. Each column of $\mathbf{W}_{\mathrm{in}}$ is normalized to unit $L_2$ norm and then multiplied by the input scaling factor $\gamma\sqrt{N/D}$ (cf.\ Algorithm~\ref{alg:deterministic-win}).

\subsection{Generalized synchronization under slow-spectrum preshaping}
\label{sec:esp-verification}

\begin{figure*}[htbp]
  \centering
  \includegraphics[width=\textwidth]{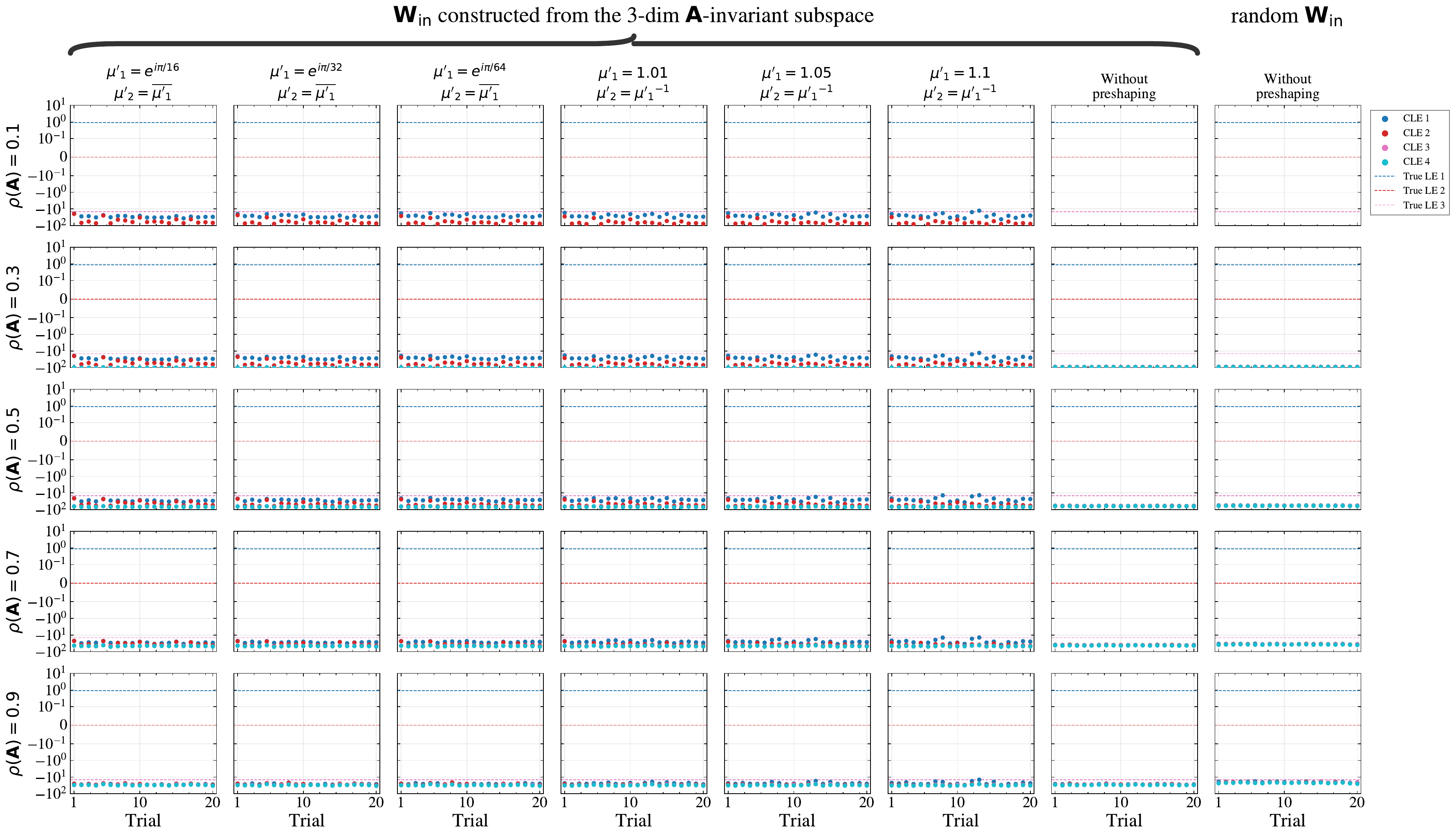}
\caption{Verification of the generalized synchronization through the maximum conditional Lyapunov exponent (CLE) of the driven reservoir, computed over 20 realizations for each configuration. Each column corresponds to a reservoir design: the six leftmost columns are the proposed method with the six eigenvalue settings of Eq.~\eqref{eq:target-eigenvalue-settings}, the second column from the right is the deterministic input layer without slow-spectrum preshaping, and the rightmost column is the standard RC with a random input layer (also without slow-spectrum preshaping). Results are shown for five initial spectral radii $\rho(\mathbf{A})$. The maximum CLE is plotted on a symmetric-log scale that is linear within $[-0.1, 0.1]$. All maximum CLEs remain negative, indicating that generalized synchronization is maintained during the listening phase.}
  \label{fig:esp-verification}
\end{figure*}

The ESP holds when the ESN update map $f$ is a uniform contraction, i.e., when the largest singular value of its state Jacobian stays below one~\cite{Jaeger2001-vs}. Our design deliberately breaks this sufficient condition, since placing slow eigenvalues with moduli close to one in $\mathbf{A}'$ can drive the largest singular value of $\mathbf{A}'_{\mathrm{eff}}$ above one.

However, this condition is only sufficient. In fact, the ESP can be formulated with respect to a particular driving input, and hence the global ESP is stronger than what is required for learning a given chaotic time series~\cite{Suetani2026-hg, Manjunath2013-bj}. It suffices that the reservoir be in generalized synchronization with the target system. GS holds when all CLEs of the reservoir are negative~\cite{Kocarev1996-mt}. We therefore calculate the CLEs from the Jacobian $\mathbf{J}_t^{\mathrm{RC}}$ of the listening-phase update Eq.~\eqref{eq:rc-update} with respect to the state $\bm{r}$,
\begin{align}
  \mathbf{J}_{t}^{\mathrm{RC}}
  &=\left. \frac{\partial f(\bm{r}, \bm{u})}{\partial \bm{r}} \right|_{\bm{r} = \bm{r}_t, \bm{u}=\bm{u}_t} \nonumber \\
  &= (1-\alpha)\mathbf{I} \nonumber \\
  &\quad+ \alpha\,
  \mathrm{diag}\!\left(1 - \bm{\tanh}^2(\mathbf{A}' \bm{r}_t +
  \mathbf{W}_{\mathrm{in}} \bm{u}_t + \bm{b})\right) \mathbf{A}'.
  \label{eq:rc-jacobian}
\end{align}

Figure~\ref{fig:esp-verification} shows the distribution of the maximum CLE computed for five initial spectral radii $\rho(\mathbf{A})$ of the reservoir transition matrix. The six leftmost columns correspond to the proposed method with the six eigenvalue settings of Eq.~\eqref{eq:target-eigenvalue-settings} and $\mathbf{W}_{\mathrm{in}}$ constructed from the associated three-dimensional real invariant subspace. The second column from the right corresponds to a randomly generated $\mathbf{A}$ with $\mathbf{W}_{\mathrm{in}}$ built from a three-dimensional real invariant subspace of $\mathbf{A}_{\mathrm{eff}}$ but without slow-spectrum preshaping (Sec.~\ref{sec:deterministic-win-limitations}), and the rightmost column corresponds to the standard construction with random $\mathbf{W}_{\mathrm{in}}$ and $\mathbf{A}$ (Sec.~\ref{sec:standard-rc-limitations}).

The maximum CLEs with slow-spectrum preshaping tend to be somewhat larger than those without it, but they remain strictly negative in all settings evaluated. This result quantitatively confirms that even when slow eigenvalues are introduced and the contractivity of the reservoir is thereby relaxed, generalized synchronization is maintained in the listening phase. Furthermore, these maximum CLEs are smaller than the negative Lyapunov exponent of the target Lorenz system (about $-14.57$), except in 7 realizations of Case 6. This suggests that the transverse attraction toward the reconstructed attractor set is stronger than the attraction intrinsic to the original attractor. A closed-loop RC built from such a strongly contracting reservoir is therefore expected to estimate the negative Lyapunov exponent accurately.

\subsection{Reproduction of the Lyapunov spectrum}
\label{sec:attractor-reconstruction-performance}

\begin{figure*}[htbp]
  \centering
  \includegraphics[width=\textwidth]{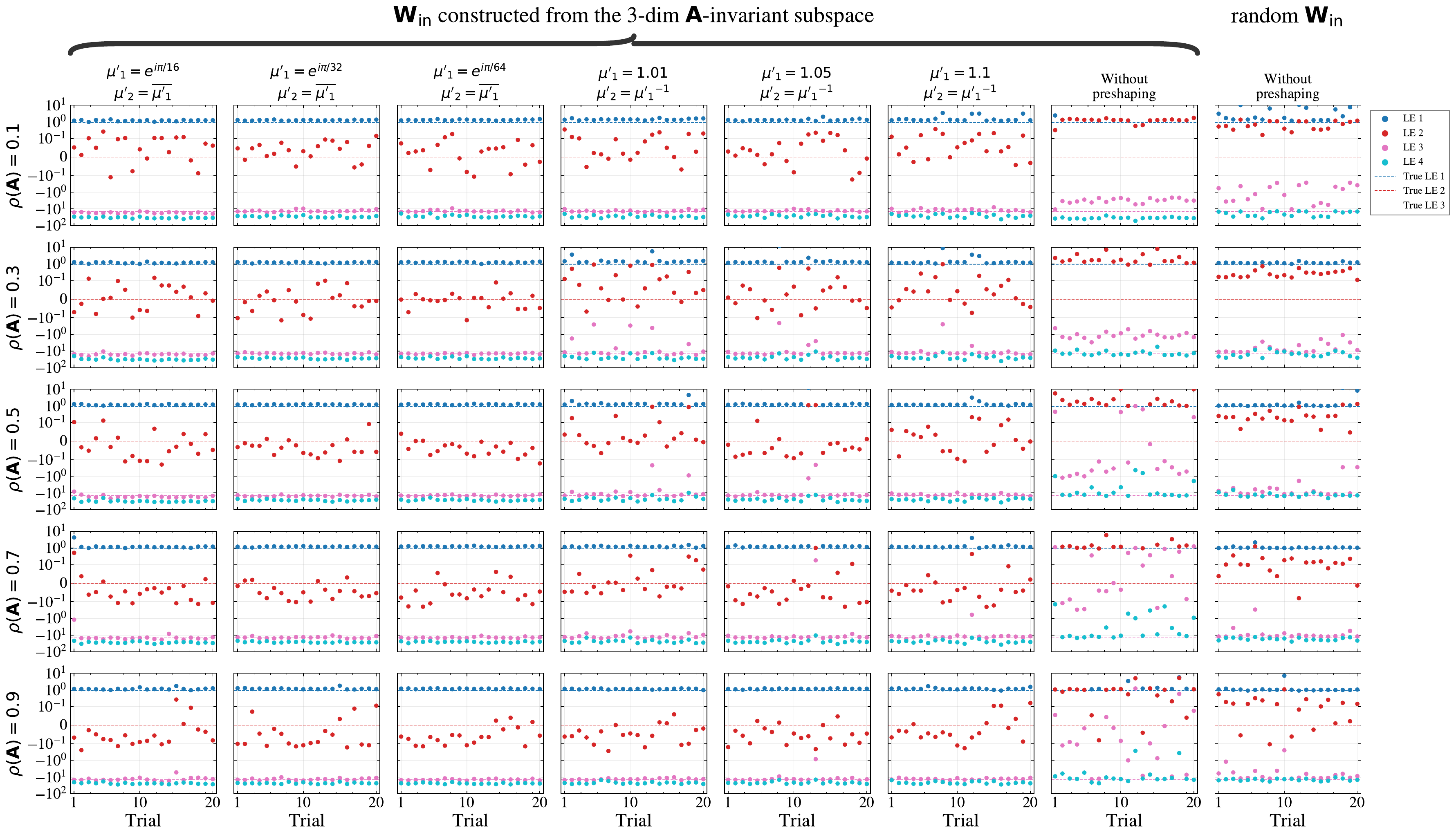}
\caption{Lyapunov spectra of the closed-loop RC, under the same reservoir designs and column layout as Fig.~\ref{fig:esp-verification}. Slow-spectrum preshaping (the six leftmost columns) reproduces the Lyapunov exponents of the original Lorenz attractor (horizontal dashed lines) and suppresses the spurious exponents. The estimated Lyapunov exponents show much smaller variability than those for the deterministic input layer without slow-spectrum preshaping (second column from the right) and for the standard RC (rightmost column). The exponents are plotted on a symmetric-log scale that is linear within $[-0.1, 0.1]$.}
  \label{fig:closed-loop-results}
\end{figure*}

This subsection analyzes in detail how slow-spectrum preshaping affects the capability to reproduce the chaotic attractor, assessing how well the Lyapunov spectrum is reproduced. The Lyapunov spectrum of each trained closed-loop RC is computed by evaluating the closed-loop Jacobian~\eqref{eq:jacobian} along the reservoir state sequence $\mathbf{R}_T$ collected during the listening phase, which lies on the reconstructed invariant set. Figure~\ref{fig:closed-loop-results} plots these spectra, in the same settings and format as Fig.~\ref{fig:esp-verification}.

Across the 600 realizations of the proposed method (six eigenvalue settings, five spectral radii, and 20 trials each), the Lyapunov exponents of the original Lorenz attractor are approximated well in most cases. The first Lyapunov exponent is estimated successfully in almost all realizations, and the second exponent falls within $[-0.1, 0.1]$ in 467 realizations (78\%). These results indicate that the proposed method preserves the long-term behavior of the target system with high probability and reconstructs the essential dynamics robustly.

In contrast, the existing methods exhibit very large trial-to-trial variance of the Lyapunov exponents even at identical parameter settings. For the standard RC, the second exponent falls within the acceptable range $[-0.1, 0.1]$ in only 16 of the 100 realizations, and in none of them at $\rho(\mathbf{A}) = 0.1$ and $0.3$. The deterministic input-layer design without slow-spectrum preshaping also fails badly at reproducing the attractor, as already seen in Sec.~\ref{sec:deterministic-win-limitations}.

Notably, the third Lyapunov exponent lies within $\pm 3$ of that of the original system in 396 of the 467 realizations for which the second exponent falls within $[-0.1, 0.1]$. This indicates not only that the transverse Lyapunov exponents are negative, i.e., that the reconstructed attractor set is transversally attracting, but also that this transverse attraction is at least as strong as the attraction intrinsic to the original attractor. Estimating negative Lyapunov exponents from data under partial observation has been considered difficult, because spurious exponents may arise between the zero and the negative exponents~\cite{Sauer1998-la}. The proposed method nevertheless succeeds in estimating the negative exponent in most realizations because, due to the fast--slow separation, the slow submanifold provides sufficiently strong transverse attraction to keep the reconstructed attractor set within it. This outcome is consistent with the observation in Sec.~\ref{sec:esp-verification} that the CLEs lie below the Lyapunov exponents of the target system in most cases.

\subsection{Robustness over hyperparameters}
\label{sec:parameter-robustness-evaluation}

\begin{figure*}[htbp]
  \centering
  \includegraphics[width=\textwidth]{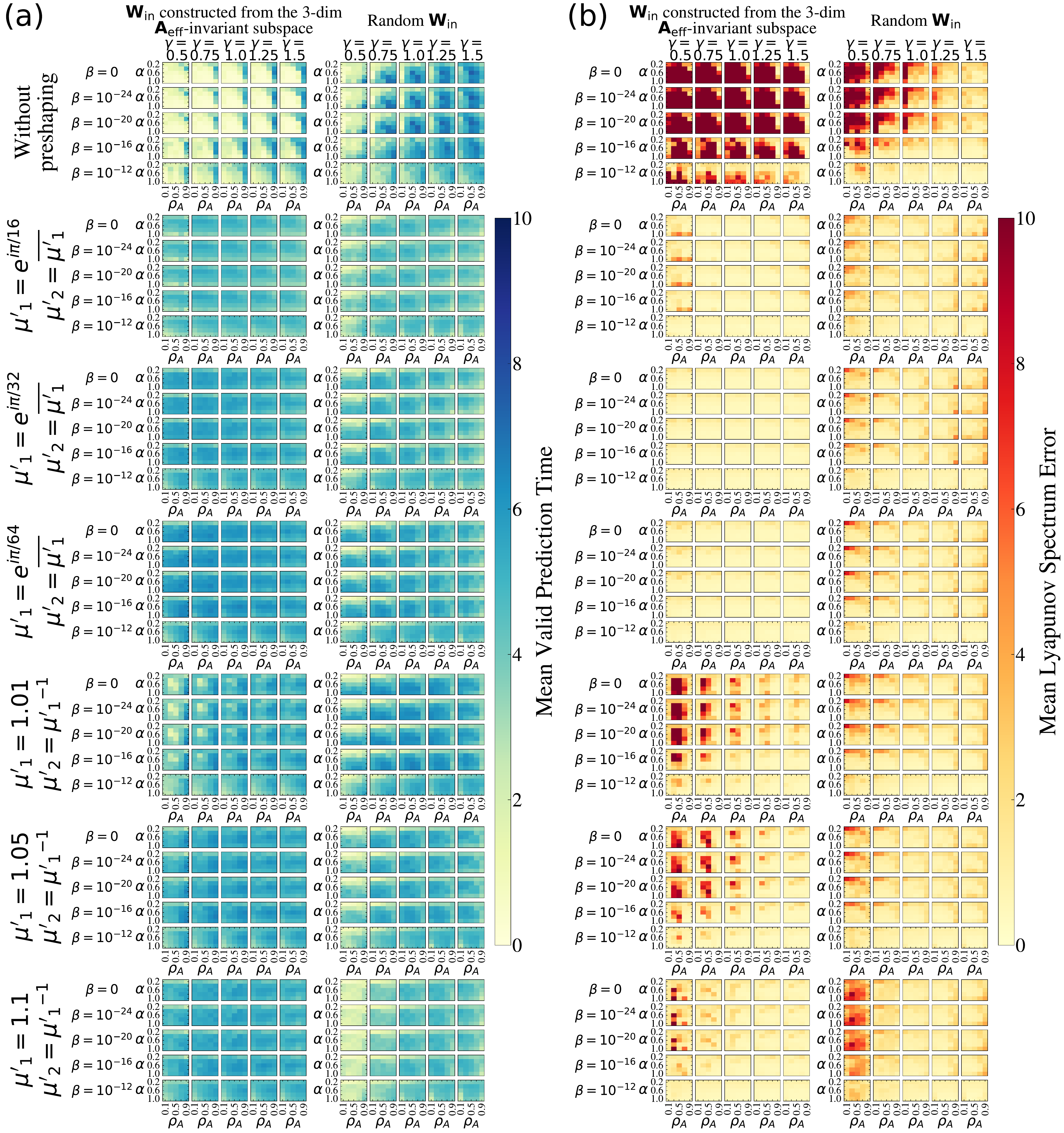}
\caption{Hyperparameter dependence of prediction performance in terms of (a) the mean valid prediction time (VPT) and (b) the mean Lyapunov spectrum error (LSE). In each small heatmap, the horizontal axis is the spectral radius $\rho$ and the vertical axis is the leaking rate $\alpha$; the small heatmaps are arranged by input scaling $\gamma$ (columns) and regularization parameter $\beta$ (rows), forming one block for each pair of reservoir-transition-matrix and input-layer designs. These blocks are further arranged by the reservoir-transition-matrix design (large rows, from top to bottom: no slow-spectrum preshaping, then Cases 1--6 of Eq.~\eqref{eq:target-eigenvalue-settings}) and by the input-layer design (large columns: deterministic on the left, random on the right). In (a), each cell reports the VPT averaged over 20 network and input-layer realizations and 50 prediction start points per realization; in (b), each cell reports the LSE averaged over the 20 realizations.}
  \label{fig:gridsearch}
\end{figure*}

To evaluate how robust the proposed method is to the choice of its hyperparameters, we performed a grid search over the ranges listed in Table~\ref{tab:experimental-parameters}: the four RC hyperparameters, namely the spectral radius $\rho$, the leaking rate $\alpha$, the input scaling $\gamma$, and the regularization parameter $\beta$, were evaluated together with the reservoir-transition-matrix design (no slow-spectrum preshaping, plus the six eigenvalue placements of Eq.~\eqref{eq:target-eigenvalue-settings}) and with the input-layer design (random, or deterministic via Algorithm~\ref{alg:deterministic-win}), giving $5^4 \times 7 \times 2 = 8750$ configurations.

Figure~\ref{fig:gridsearch} presents these results using a systematic arrangement of heatmaps that collectively represent the six-dimensional parameter space.

In Fig.~\ref{fig:gridsearch}(a), the performance of each configuration is measured by the mean VPT [Eq.~\eqref{eq:vpt}]. Because prediction performance depends on when prediction starts, we set 50 prediction start points equally spaced over the 5000 steps following the end of the training phase at $T_\mathrm{w} + T$. Up to each start point the state is updated in the open-loop form of Eq.~\eqref{eq:rc-update}, and the subsequent prediction is run in closed loop. For each configuration we further use 20 realizations of the reservoir transition matrix $\mathbf{A}$ and the input layer $\mathbf{W}_{\mathrm{in}}$. The mean VPT of each cell is therefore an average over 20 network and input-layer realizations and 50 prediction start points per realization.

In Fig.~\ref{fig:gridsearch}(b), to quantify how well the trained closed-loop RC reproduces the long-term dynamics, we introduce the \textit{Lyapunov spectrum error} (LSE),
\begin{equation}
  \mathrm{LSE}
  = \sum_{\lambda_i \ge 0} \left| \lambda_i - \hat{\lambda}_i \right|
  + \sum_{\substack{\lambda_i < 0\\ i \le \mathrm{dim}(\mathscr{M})}}
    \frac{\left| \lambda_i - \hat{\lambda}_i \right|}{\left| \lambda_i \right|},
  \label{eq:lse}
\end{equation}
where $\lambda_i$ and $\hat{\lambda}_i$ are the Lyapunov exponents of the target system and of the closed-loop RC, respectively, arranged in descending order. The non-negative exponents are compared by absolute error and the negative exponents by relative error, because faithfully reproducing the non-negative part of the spectrum is more important for capturing the dynamics than reproducing the strongly contracting negative exponent exactly; the relative error also prevents the large magnitude of the negative exponent from dominating the metric. As in Sec.~\ref{sec:attractor-reconstruction-performance}, the RC Lyapunov spectrum is computed from the listening-phase reservoir state sequence, and the LSE is evaluated against the Lorenz attractor Lyapunov exponents $(\lambda_1, \lambda_2, \lambda_3) = (0.91, 0, -14.57)$.

The standard RC, with neither slow-spectrum preshaping nor a deterministic input layer, corresponds to the top-right block of both panels. There, only 329 of the 625 parameter settings (about 53\%) achieve $\mathrm{VPT} > 4$, corresponding to the high-performing cases observed in Sec.~\ref{sec:slow-spectrum-preshaping-results} [Fig.~\ref{fig:gridsearch}(a)], indicating that the prediction performance depends strongly on the combination of $\rho$, $\alpha$, $\gamma$, and $\beta$. The LSE is likewise very large where the input scaling $\gamma$ and the regularization $\beta$ are small [Fig.~\ref{fig:gridsearch}(b)], indicating poor reproduction of the Lyapunov spectrum. This confirms that the capability to reproduce the long-term dynamics is also strongly hyperparameter-dependent. 

The proposed method, which combines slow-spectrum preshaping with the deterministic input layer, corresponds to the six blocks in the left column below the top row (Cases 1--6). These settings attain $\mathrm{VPT} > 4$ in 414, 576, 579, 457, 556, and 560 of the 625 parameter settings, respectively, exceeding the standard RC for every eigenvalue placement. In the preliminary experiments of Secs.~\ref{sec:standard-rc-limitations} and \ref{sec:slow-spectrum-preshaping-results}, the mean VPT of the standard RC exceeded that of the proposed method, but this appears to be accidental rather than a general trend across the hyperparameter space. Turning to the mean LSE [Fig.~\ref{fig:gridsearch}(b)], the rotation-like Cases 1--3 keep it low over almost the entire parameter region, indicating successful reproduction of the dynamics. For the saddle-like Cases 4--6, by contrast, it becomes large over part of the parameter region. Among these saddle-like placements, however, the LSE decreases as the separation between $\hat{\mu}_1$ and $\hat{\mu}_2$ increases. For the present setting (observation of the Lorenz $x$ variable), preshaping the rotation-like eigenvalues of Cases 2 and 3 achieves highly robust reproduction of the dynamics. We note, however, that these eigenvalue placements were chosen and explored heuristically, and the optimal placement may differ depending on the problem setting; the dependence on the target system and the observed variable is investigated in Appendix~\ref{sec:dysts-validation}.

In contrast, when only the input layer is designed deterministically without slow-spectrum preshaping (top-left block), the prediction performance deteriorates over almost the entire parameter region, as observed in Sec.~\ref{sec:deterministic-win-limitations}. When slow-spectrum preshaping is applied but the input layer is constructed randomly (the six blocks in the right column below the top row), regions of high performance do appear, but the dependence on the hyperparameters remains strong. Introducing slow eigenvalues into the reservoir transition matrix is therefore not sufficient by itself. Constructing the input layer deterministically from the slow spectral subspace is essential for keeping the reconstructed trajectories in the attracting slow submanifold over a wide parameter range.

We note that the longest mean VPT recorded in this grid search was attained by a pinpoint-optimized standard RC, which reached a VPT of about 7.50 at $\alpha = 0.6$, $\rho = 0.5$, $\beta = 0$, and $\gamma = 1.5$. With sufficient search, standard RC can thus exhibit very high short-term prediction performance. Such high-performance regions are narrow, however, and identifying them in advance for an unknown target system is not easy. The claim of this section is therefore not that the proposed method always outperforms the well-tuned standard RC, but that it delivers consistently good prediction over a wide hyperparameter region and thereby greatly reduces the reliance on \textit{a posteriori} parameter search.

\section{Conclusions and discussion}
\label{sec:conclusion-and-discussion}

Attractor Reconstruction in Attracting Submanifolds (ARAS) provides an approach to robustly reproducing target dynamics by placing the reconstructed invariant set within a transversally attracting submanifold of the internal state space. We realize ARAS for RC under partial observation by combining slow-spectrum preshaping with the deterministic input-layer design of Ref.~\onlinecite{Oishi2026-fz}. Slow-spectrum preshaping introduces a small number of slow eigenvalues into the reservoir transition matrix, inducing slow modes that retain a memory of past inputs. The deterministic input-layer design then makes the slow submanifold formed by these modes transversally attracting by keeping the remaining modes fast-contracting, so that the target attractor can be reconstructed within the attracting slow submanifold. Unlike the common heuristic of enhancing reservoir memory by increasing the spectral radius or decreasing the leaking rate, which can introduce excess slow modes, the proposed construction preserves the fast--slow separation and thereby suppresses transverse instabilities. Our numerical experiments showed that this design reproduces and predicts various chaotic dynamics robustly (see also Appendix~\ref{sec:dysts-validation}). The driven reservoir achieves generalized synchronization, the closed-loop RC reproduces the Lyapunov spectrum of the Lorenz system in most realizations, and the prediction performance remains high over a far wider hyperparameter region than for standard RC.

The proposed construction is \textit{a priori}---the reservoir and input layer are constructed before the listening phase and thus task agnostic, in contrast to \textit{a posteriori} hyperparameter search---but some design parameters within our approach are still selected manually:
\begin{itemize}
\item How to select the number $k$ of slow eigenvalues and the dimension $D'$ of the invariant subspace remains to be investigated. Existing methods \cite{Kennel1992-rz, Liebert1991-dz, Pecora2007-fj} for estimating the embedding dimension may be applicable here.
\item How to select the placement of the specified eigenvalues also remains to be investigated. We confirmed that a particular rotation-like placement is effective for the Lorenz system under observation of the $x$ variable, but the optimal placement may differ for other dynamical systems and observables, as the benchmark results of Appendix~\ref{sec:dysts-validation} indicate. Methods \cite{Gao1993-fj, Kantz2004-wo} that estimate the characteristic timescale of a system, or that search for the optimal delay in delay-coordinate embedding, may enable a data-driven choice of the placement.
\end{itemize}

Furthermore, preshaping the spectrum of the reservoir transition matrix is not the only route to achieving ARAS \textit{a priori}, and exploring other routes is an important direction for future research. Beyond RC, incorporating the ARAS perspective into the design of other data-driven learners of dynamical systems is also a promising direction.

\begin{acknowledgments}
We thank Taiki Yamada for fruitful discussions.
This work was supported by JSPS KAKENHI Grant Number JP25KJ1714, JST CREST Grant No. JPMJCR25R1, JST ALCA-Next Grant No. JPMJAN23F2, JST Moonshot R\&D Grant No. JPMJMS2021, JST PRESTO Grant No. JPMJPR25K5 and a project, JPNP14004, commissioned by the New Energy and Industrial Technology Development Organization (NEDO).
\end{acknowledgments}

\section*{Author declarations}
\subsection*{Conflict of Interest}
The authors have no conflicts to disclose.

\subsection*{Author Contributions}
Satoshi Oishi: Conceptualization (lead); Data curation (lead); Formal analysis (lead); Funding acquisition (equal); Investigation (lead); Methodology (lead); Project administration (lead); Resources (equal); Software (lead); Validation (lead); Writing - original draft (lead); Writing - review \& editing (equal).
Hiroshi Yamashita: Conceptualization (supporting); Formal analysis (equal); Supervision (equal); Writing - review \& editing (equal).
Hideyuki Suzuki: Funding acquisition (lead); Resources (equal); Writing - original draft (supporting); Writing - review \& editing (supporting).
Sho Shirasaka: Conceptualization (equal); Formal analysis (equal); Funding acquisition (equal); Investigation (equal); Supervision (lead); Writing - original draft (supporting); Writing - review \& editing (lead).

\section*{Data availability}
The implementation used in this study is available on GitHub:
\url{https://github.com/nonLinMathSciLab/slow-spectrum-preshaping}

\appendix

\begin{figure*}[htbp]
  \centering
  \includegraphics[width=\textwidth]{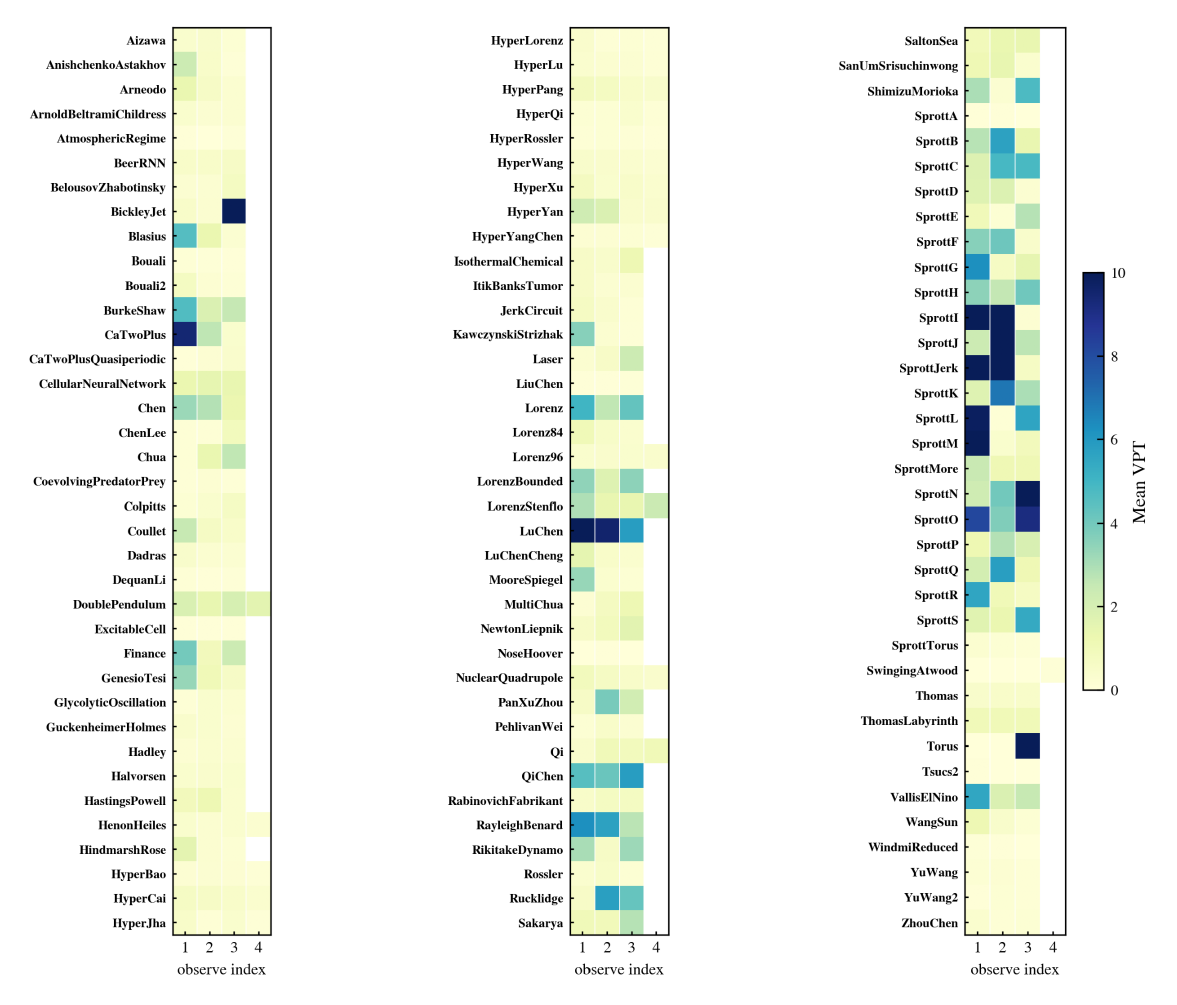}
\caption{Prediction performance of the proposed method on the Dysts benchmark of low-dimensional chaotic systems~\cite{Gilpin2021-wp, Gilpin2023-ua}. Each row corresponds to an autonomous, delay-free system from the benchmark, and each column (observe index) corresponds to the single state variable supplied as the partial observation. Three-dimensional systems use indices $1$--$3$, and four-dimensional systems use indices $1$--$4$. The color of each cell encodes the valid prediction time averaged over 20 realizations and 50 prediction start points per realization}
  \label{fig:dysts-validation}
\end{figure*}

\section{Lorenz-63 system}
\label{sec:lorenz-equation}

The Lorenz-63 system~\cite{Lorenz1963-kf} is a three-dimensional continuous-time dynamical system defined by
\begin{equation}
\begin{aligned}
    \dot{x} &= \sigma (y - x), \\
    \dot{y} &= x (\rho - z) - y, \\
    \dot{z} &= xy - \beta z .
\end{aligned}
\end{equation}
Throughout this paper, we use the standard parameter values $\sigma = 10$, $\rho = 28$, and $\beta = 8/3$.

The time-series data were generated following the procedure used in the Dysts dataset~\cite{Gilpin2021-wp,Gilpin2023-ua}.
Trajectories were aligned by their dominant Fourier timescale and sampled at 100 time points per dominant period. For the Lorenz-63 system, this corresponds to a sampling interval of $\varDelta t = 0.015008$.

\section{Validation on a benchmark suite of chaotic systems}
\label{sec:dysts-validation}

We further examine the performance of the proposed method on Dysts, a dataset of low-dimensional chaotic dynamical systems~\cite{Gilpin2021-wp, Gilpin2023-ua}. From this dataset we select the autonomous systems without delay, and we train and predict each observed variable with the proposed method (Fig.~\ref{fig:dysts-validation}). For three-dimensional systems such as the Lorenz system, the introduced slow eigenvalues are $\hat{\mu}_1 = e^{\mathrm{i}\pi/32}$ and $\hat{\mu}_2 = \bar{\hat{\mu}}_1$. For four-dimensional chaotic systems, we introduce the three slow eigenvalues $\hat{\mu}_1 = e^{\mathrm{i}\pi/32}$, $\hat{\mu}_2 = \bar{\hat{\mu}}_1$, and $\hat{\mu}_3 = 1$ (i.e., $k = 3$) and use a four-dimensional invariant subspace for the input layer ($D' = 4$). The other experimental conditions follow those of Sec.~\ref{sec:slow-spectrum-preshaping-results}, except that the spectral radius is set to $\rho(\mathbf{A}) = 0.3$.

The results show high prediction accuracy for several systems, but not for all of them. Moreover, for the same system, the prediction can succeed or fail depending on which variable is observed. These observations indicate that, while the proposed method provides a framework for attractor reconstruction under partial observation, the optimal choice of the number and placement of the slow eigenvalues for a given target system and observable remains an important open problem.

\section{Delay-coordinate feedback for autonomous prediction}
\label{sec:delay-coordinates}

This appendix demonstrates that feeding the RC with the attractor reconstructed from delay coordinates does not ensure reproduction of the target dynamics, even with the deterministic input-layer design of Ref.~\onlinecite{Oishi2026-fz}. We compare two delay-feedback schemes used in prior work, each with the deterministic input layer and with a random input layer.

\subsection{Setup and delay-coordinate embedding}
\label{sec:delay-setup}

We use the same RC settings as in Secs.~\ref{sec:standard-rc-limitations} and \ref{sec:deterministic-win-limitations} (the Figs.~\ref{fig:standard-rc-results}--\ref{fig:slow-spectrum-preshaping-results} column of Table~\ref{tab:experimental-parameters}). The target is the Lorenz-63 system, of which only the normalized scalar observation $u_t \in \mathbb{R}$ (the $x$ variable) is available. We embed this observation into delay coordinates, choosing the embedding dimension and delays with the delay-optimization method PECUZAL~\cite{Kraemer2021-ln}. Here we adopt
\begin{equation}
\bm{q}_t = \begin{bmatrix} u_t & u_{t-5} & u_{t-9} \end{bmatrix}^{\top} \in \mathbb{R}^3,
\end{equation}
so that the scalar observation is mapped to a three-dimensional delay vector $\bm{q}_t$, which serves as the RC input. During the listening phase, the reservoir state evolves as in Eq.~\eqref{eq:rc-update}, here with $\alpha = 1$ and the delay vector $\bm{q}_t$ as input:
\begin{equation}
\bm{r}_{t+1} = \bm{\tanh}\!\left( \mathbf{A} \bm{r}_t + \mathbf{W}_{\mathrm{in}} \bm{q}_t + \bm{b} \right).
\end{equation}
In the prediction phase, $\bm{q}_t$ is generated from the RC's own output instead of the observed sequence, closing the loop.

\begin{figure}[htbp]
  \centering
  \includegraphics[width=\columnwidth]{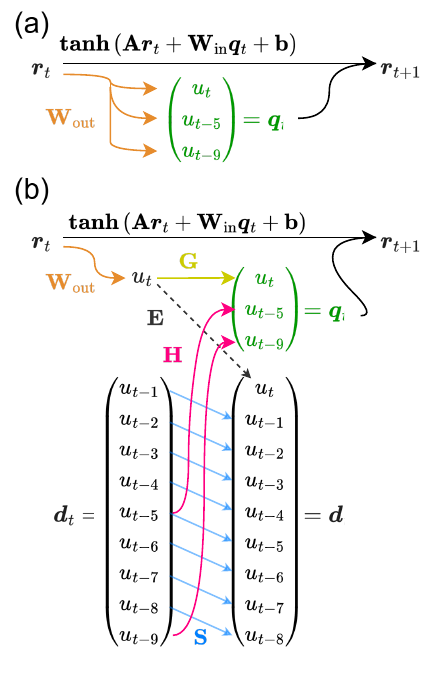}
\caption{The two delay-coordinate feedback schemes. (a) \texttt{output=input}: the output layer generates the entire delay vector $\bm{q}_t$. (b) \texttt{output=observed}: it generates only the next value $u_t$ and keeps the past values as a history $\bm{d}_t$ that is stored and shifted.}
  \label{fig:delay-schemes}
\end{figure}

\subsection{Taxonomy of delay feedback}
\label{sec:delay-taxonomy}

Prior RC/ESN studies on prediction under partial observation can be organized by where and how the delay structure is introduced. The works cited here fall into three types.

\emph{(i) Output and feed back the delay vector of the observation} (\texttt{output=input}): the delay vector is constructed from the partial observation and used as the input, and the output layer produces the entire delay vector and returns it as the next input~\cite{Nakai2018-cw, Gottwald2021-zx} [Fig.~\ref{fig:delay-schemes}(a)].

\emph{(ii) Output only the next observation and shift the observation history} (\texttt{output=observed}): the output layer produces only the next observation, and the past components of the delay vector are assembled from an observation-history buffer (a shift register) that stores previously generated values~\cite{Goswami2023-th} [Fig.~\ref{fig:delay-schemes}(b)].

\emph{(iii) Time-shift the reservoir internal state}: a time delay is applied to the reservoir states rather than to the observation, and past reservoir states serve as additional readout features~\cite{Duan2023-ge, Jaurigue2025-wy}.

\subsection{Two feedback schemes}
\label{sec:delay-schemes}
We now formulate schemes (i) and (ii) as autonomous dynamical systems [Fig.~\ref{fig:delay-schemes}(a) and (b)]. Both share the listening-phase update above and differ in how the delay vector $\bm{q}_t$ is generated in the prediction phase, and hence in the state of the resulting autonomous system.

\emph{(a) \texttt{output=input}.} The output layer produces the entire delay vector ($\mathbf{W}_{\mathrm{out}}^{\mathrm{in}} \in \mathbb{R}^{3 \times N}$), and the prediction phase reads
\begin{equation}
\bm{q}_t = \mathbf{W}_{\mathrm{out}}^{\mathrm{in}} \bm{r}_t, \qquad
\bm{r}_{t+1} = \bm{\tanh}\!\left( \left( \mathbf{A} + \mathbf{W}_{\mathrm{in}} \mathbf{W}_{\mathrm{out}}^{\mathrm{in}} \right) \bm{r}_t + \bm{b} \right),
\end{equation}
so the autonomous state is $\bm{r}_t \in \mathbb{R}^N$ alone. The past components $u_{t-5}$ and $u_{t-9}$ of $\bm{q}_t$ are re-estimated from $\bm{r}_t$ at every step. Nothing forces them to equal the values the RC itself generated five and nine steps earlier, so the delay structure of the input is not enforced in closed loop.

\emph{(b) \texttt{output=observed}.} The output layer produces only the next value ($\mathbf{W}_{\mathrm{out}}^{\mathrm{obs}} \in \mathbb{R}^{1 \times N}$), $u_t = \mathbf{W}_{\mathrm{out}}^{\mathrm{obs}} \bm{r}_t$. Past values are not re-predicted but taken from a history vector that stores the values generated at earlier steps,
\begin{equation}
\bm{d}_t = \begin{bmatrix} u_{t-1} & u_{t-2} & \cdots & u_{t-9} \end{bmatrix}^{\top} \in \mathbb{R}^9.
\end{equation}
The delay input is assembled from $u_t$ and the history, and the history is shifted one step backward:
\begin{equation}
\bm{q}_t = \mathbf{G} u_t + \mathbf{H} \bm{d}_t, \qquad
\bm{d}_{t+1} = \mathbf{E} u_t + \mathbf{S} \bm{d}_t.
\end{equation}
Here the constant matrices $\mathbf{G}$, $\mathbf{H}$, $\mathbf{E}$, and $\mathbf{S}$ are
\begin{widetext}
\begin{equation}
\begin{gathered}
\mathbf{G} = \begin{bmatrix} 1 \\ 0 \\ 0 \end{bmatrix} \in \mathbb{R}^{3 \times 1}, \qquad
\mathbf{H} = \begin{bmatrix} 0&0&0&0&0&0&0&0&0 \\ 0&0&0&0&1&0&0&0&0 \\ 0&0&0&0&0&0&0&0&1 \end{bmatrix} \in \mathbb{R}^{3 \times 9}, \\[6pt]
\mathbf{E} = \begin{bmatrix} 1 \\ 0 \\ \vdots \\ 0 \end{bmatrix} \in \mathbb{R}^{9 \times 1}, \qquad
\mathbf{S} = \begin{bmatrix} 0&0&\cdots&0&0 \\ 1&0&\cdots&0&0 \\ 0&1&\cdots&0&0 \\ \vdots& &\ddots& &\vdots \\ 0&0&\cdots&1&0 \end{bmatrix} \in \mathbb{R}^{9 \times 9}.
\end{gathered}
\end{equation}
\end{widetext}
Thus $\mathbf{G}$ injects $u_t$ into the current slot, $\mathbf{H}$ selects the lag-5 and lag-9 history components, $\mathbf{E}$ inserts $u_t$ at the head of the history, and $\mathbf{S}$ shifts the whole history one step backward. The autonomous state is enlarged to include the history, $\bm{z}_t = [\, \bm{r}_t;\, \bm{d}_t \,] \in \mathbb{R}^{N+9}$, and its autonomous map is
\begin{widetext}
\begin{equation}
\bm{z}_{t+1} = \hat{f}_{\mathrm{obs}}(\bm{z}_t) := \begin{bmatrix}
\bm{\tanh}\!\left( \left( \mathbf{A} + \mathbf{W}_{\mathrm{in}} \mathbf{G} \mathbf{W}_{\mathrm{out}}^{\mathrm{obs}} \right) \bm{r}_t + \mathbf{W}_{\mathrm{in}} \mathbf{H} \bm{d}_t + \bm{b} \right) \\[4pt]
\mathbf{E} \mathbf{W}_{\mathrm{out}}^{\mathrm{obs}} \bm{r}_t + \mathbf{S} \bm{d}_t
\end{bmatrix}.
\end{equation}
\end{widetext}
Its linear part, \begin{equation} \mathbf{W}_{\mathrm{cl}}^{\mathrm{obs}} = \begin{bmatrix} \mathbf{A} + \mathbf{W}_{\mathrm{in}} \mathbf{G} \mathbf{W}_{\mathrm{out}}^{\mathrm{obs}} & \mathbf{W}_{\mathrm{in}} \mathbf{H} \\ \mathbf{E} \mathbf{W}_{\mathrm{out}}^{\mathrm{obs}} & \mathbf{S} \end{bmatrix} \in \mathbb{R}^{(N+9) \times (N+9)}, \end{equation} is the closed-loop matrix whose eigenvalues are shown in Fig.~\ref{fig:delay-results-fig}(b) and (d).

Both schemes train the output layer by linear regression from the same reservoir state sequence $\{\bm{r}_t\}$, but differ in the training target (the entire delay vector $\bm{q}_t$ for scheme (a), only the next value $u_t$ for scheme (b)), the readout dimension, and how $\bm{q}_t$ is generated during autonomous prediction. In addition, for the input layer $\mathbf{W}_{\mathrm{in}}$ we compare the deterministic input-layer design of Ref.~\onlinecite{Oishi2026-fz} and a random input layer.

\subsection{Results}
\label{sec:delay-results}

\begin{figure*}[htbp]
  \centering
  \includegraphics[width=\textwidth]{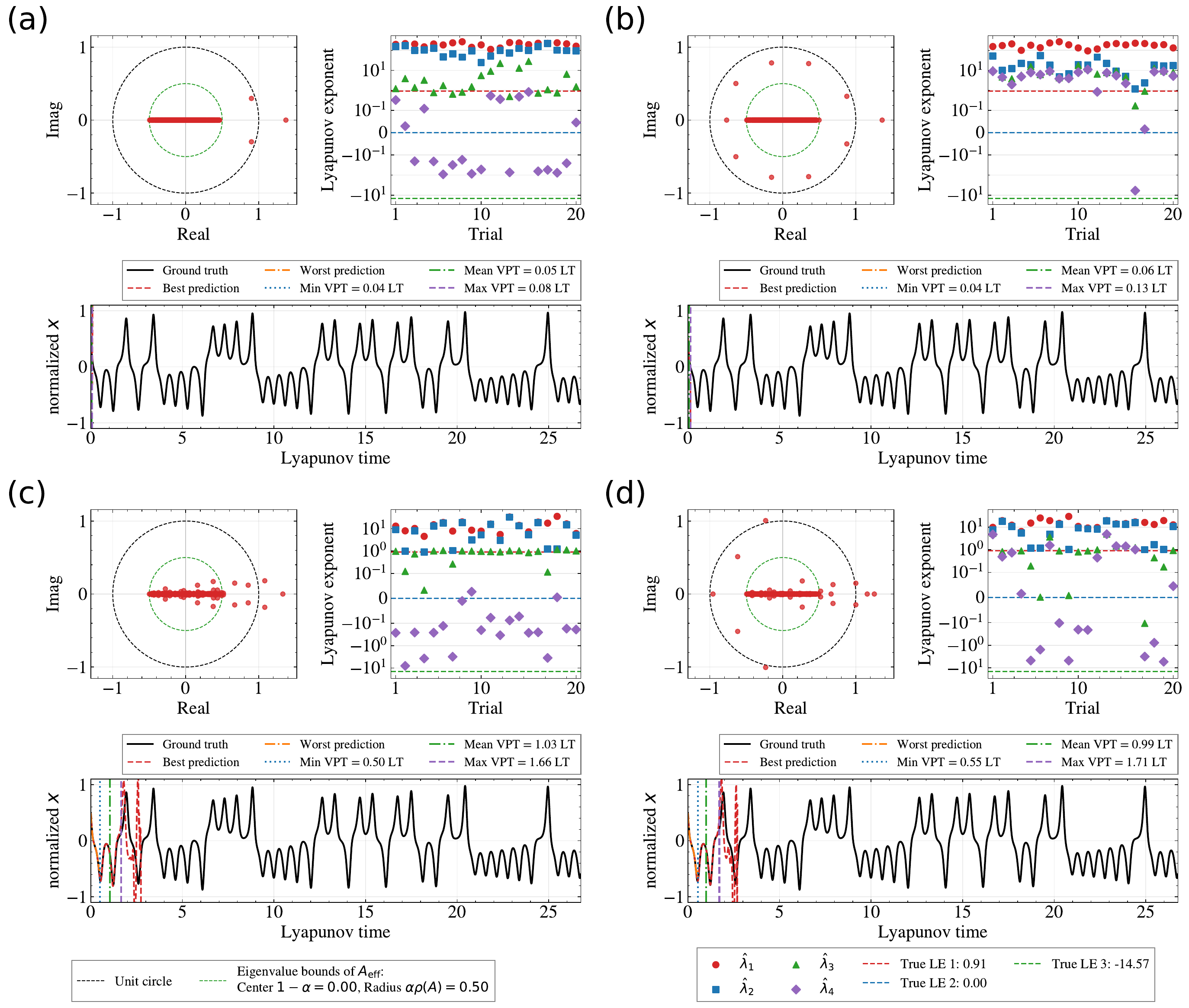}
\caption{Delay-coordinate feedback on the Lorenz-63 system for the four conditions: (a)~deterministic $\mathbf{W}_{\mathrm{in}}$, \texttt{output=input}; (b)~deterministic $\mathbf{W}_{\mathrm{in}}$, \texttt{output=observed}; (c)~random $\mathbf{W}_{\mathrm{in}}$, \texttt{output=input}; (d)~random $\mathbf{W}_{\mathrm{in}}$, \texttt{output=observed}. In each panel: eigenvalue distribution of the closed-loop matrix (top left), estimated Lyapunov spectrum over the 20 realizations with the original Lorenz exponents as dashed lines (top right), and observed-variable prediction overlay with the valid prediction time (bottom).}
  \label{fig:delay-results-fig}
\end{figure*}

We ran autonomous prediction and Lyapunov-spectrum estimation for the four conditions (input-layer construction) $\times$ (feedback scheme), with 20 independent realizations each. Figure~\ref{fig:delay-results-fig} shows the results for the four conditions. The closed-loop matrix is $\mathbf{A} + \mathbf{W}_{\mathrm{in}} \mathbf{W}_{\mathrm{out}}^{\mathrm{in}}$ for \texttt{output=input} and $\mathbf{W}_{\mathrm{cl}}^{\mathrm{obs}}$ for \texttt{output=observed}. In every condition the prediction diverges within a few Lyapunov times and the estimated Lyapunov spectrum departs from the original Lorenz exponents.

\subsection{Summary}
\label{sec:delay-summary}

None of the conditions achieved stable short-term prediction or long-term reproduction of the dynamics. That is, even when the attractor is reconstructed by delay coordinates, this alone does not guarantee its reproduction under autonomous prediction. Because we did not exhaustively search over the delays or the RC hyperparameters, tuning them might reach a prediction performance comparable to that reported in prior work~\cite{Nakai2018-cw, Gottwald2021-zx, Goswami2023-th, Duan2023-ge, Jaurigue2025-wy}. The key point, however, is that delay-coordinate feedback by itself does not guarantee high performance without such tuning. Robust reproduction of the dynamics requires a mechanism that reconstructs the attractor in the state space of the autonomous closed-loop model and keeps the reconstructed set transversally attracting, which is what ARAS provides.

\newpage

\bibliography{paperpile}

\begin{thebibliography}{73}%
\makeatletter
\providecommand \@ifxundefined [1]{%
 \@ifx{#1\undefined}
}%
\providecommand \@ifnum [1]{%
 \ifnum #1\expandafter \@firstoftwo
 \else \expandafter \@secondoftwo
 \fi
}%
\providecommand \@ifx [1]{%
 \ifx #1\expandafter \@firstoftwo
 \else \expandafter \@secondoftwo
 \fi
}%
\providecommand \natexlab [1]{#1}%
\providecommand \enquote  [1]{``#1''}%
\providecommand \bibnamefont  [1]{#1}%
\providecommand \bibfnamefont [1]{#1}%
\providecommand \citenamefont [1]{#1}%
\providecommand \href@noop [0]{\@secondoftwo}%
\providecommand \href [0]{\begingroup \@sanitize@url \@href}%
\providecommand \@href[1]{\@@startlink{#1}\@@href}%
\providecommand \@@href[1]{\endgroup#1\@@endlink}%
\providecommand \@sanitize@url [0]{\catcode `\\12\catcode `\$12\catcode `\&12\catcode `\#12\catcode `\^12\catcode `\_12\catcode `\%12\relax}%
\providecommand \@@startlink[1]{}%
\providecommand \@@endlink[0]{}%
\providecommand \url  [0]{\begingroup\@sanitize@url \@url }%
\providecommand \@url [1]{\endgroup\@href {#1}{\urlprefix }}%
\providecommand \urlprefix  [0]{URL }%
\providecommand \Eprint [0]{\href }%
\providecommand \doibase [0]{https://doi.org/}%
\providecommand \selectlanguage [0]{\@gobble}%
\providecommand \bibinfo  [0]{\@secondoftwo}%
\providecommand \bibfield  [0]{\@secondoftwo}%
\providecommand \translation [1]{[#1]}%
\providecommand \BibitemOpen [0]{}%
\providecommand \bibitemStop [0]{}%
\providecommand \bibitemNoStop [0]{.\EOS\space}%
\providecommand \EOS [0]{\spacefactor3000\relax}%
\providecommand \BibitemShut  [1]{\csname bibitem#1\endcsname}%
\let\auto@bib@innerbib\@empty
\bibitem [{\citenamefont {Brunton}\ and\ \citenamefont {Kutz}(2022)}]{Brunton2022-bj}%
  \BibitemOpen
  \bibfield  {author} {\bibinfo {author} {\bibfnamefont {S.~L.}\ \bibnamefont {Brunton}}\ and\ \bibinfo {author} {\bibfnamefont {J.~N.}\ \bibnamefont {Kutz}},\ }\href@noop {} {\emph {\bibinfo {title} {Data-driven science and engineering: Machine learning, dynamical systems, and control}}}\ (\bibinfo  {publisher} {Cambridge University Press},\ \bibinfo {year} {2022})\BibitemShut {NoStop}%
\bibitem [{\citenamefont {Kantz}\ and\ \citenamefont {Schreiber}(2004)}]{Kantz2004-wo}%
  \BibitemOpen
  \bibfield  {author} {\bibinfo {author} {\bibfnamefont {H.}~\bibnamefont {Kantz}}\ and\ \bibinfo {author} {\bibfnamefont {T.}~\bibnamefont {Schreiber}},\ }\href@noop {} {{\selectlanguage {en}\emph {\bibinfo {title} {Nonlinear Time Series Analysis}}}}\ (\bibinfo  {publisher} {Cambridge University Press},\ \bibinfo {year} {2004})\BibitemShut {NoStop}%
\bibitem [{\citenamefont {Özalp}, \citenamefont {Margazoglou},\ and\ \citenamefont {Magri}(2023)}]{Ozalp2023-do}%
  \BibitemOpen
  \bibfield  {author} {\bibinfo {author} {\bibfnamefont {E.}~\bibnamefont {Özalp}}, \bibinfo {author} {\bibfnamefont {G.}~\bibnamefont {Margazoglou}},\ and\ \bibinfo {author} {\bibfnamefont {L.}~\bibnamefont {Magri}},\ }\bibfield  {title} {{\selectlanguage {en}\enquote {\bibinfo {title} {Reconstruction, forecasting, and stability of chaotic dynamics from partial data},}\ }}\href@noop {} {\bibfield  {journal} {\bibinfo  {journal} {Chaos}\ }\textbf {\bibinfo {volume} {33}},\ \bibinfo {pages} {093107} (\bibinfo {year} {2023})}\BibitemShut {NoStop}%
\bibitem [{\citenamefont {Packard}\ \emph {et~al.}(1980)\citenamefont {Packard}, \citenamefont {Crutchfield}, \citenamefont {Farmer},\ and\ \citenamefont {Shaw}}]{Packard1980-ay}%
  \BibitemOpen
  \bibfield  {author} {\bibinfo {author} {\bibfnamefont {N.~H.}\ \bibnamefont {Packard}}, \bibinfo {author} {\bibfnamefont {J.~P.}\ \bibnamefont {Crutchfield}}, \bibinfo {author} {\bibfnamefont {J.~D.}\ \bibnamefont {Farmer}},\ and\ \bibinfo {author} {\bibfnamefont {R.~S.}\ \bibnamefont {Shaw}},\ }\bibfield  {title} {{\selectlanguage {en}\enquote {\bibinfo {title} {Geometry from a time series},}\ }}\href@noop {} {\bibfield  {journal} {\bibinfo  {journal} {Phys. Rev. Lett.}\ }\textbf {\bibinfo {volume} {45}},\ \bibinfo {pages} {712--716} (\bibinfo {year} {1980})}\BibitemShut {NoStop}%
\bibitem [{\citenamefont {Takens}(1981)}]{Takens1981-mw}%
  \BibitemOpen
  \bibfield  {author} {\bibinfo {author} {\bibfnamefont {F.}~\bibnamefont {Takens}},\ }\bibfield  {title} {\enquote {\bibinfo {title} {Detecting strange attractors in turbulence},}\ }in\ \href@noop {} {\emph {\bibinfo {booktitle} {Dynamical Systems and Turbulence, Warwick 1980}}},\ \bibinfo {editor} {edited by\ \bibinfo {editor} {\bibfnamefont {D.}~\bibnamefont {Rand}}\ and\ \bibinfo {editor} {\bibfnamefont {L.-S.}\ \bibnamefont {Young}}}\ (\bibinfo  {publisher} {Springer Berlin Heidelberg},\ \bibinfo {address} {Berlin, Heidelberg},\ \bibinfo {year} {1981})\ pp.\ \bibinfo {pages} {366--381}\BibitemShut {NoStop}%
\bibitem [{\citenamefont {Sauer}, \citenamefont {Yorke},\ and\ \citenamefont {Casdagli}(1991)}]{Sauer1991-ew}%
  \BibitemOpen
  \bibfield  {author} {\bibinfo {author} {\bibfnamefont {T.}~\bibnamefont {Sauer}}, \bibinfo {author} {\bibfnamefont {J.~A.}\ \bibnamefont {Yorke}},\ and\ \bibinfo {author} {\bibfnamefont {M.}~\bibnamefont {Casdagli}},\ }\bibfield  {title} {\enquote {\bibinfo {title} {Embedology},}\ }\href@noop {} {\bibfield  {journal} {\bibinfo  {journal} {J. Stat. Phys.}\ }\textbf {\bibinfo {volume} {65}},\ \bibinfo {pages} {579--616} (\bibinfo {year} {1991})}\BibitemShut {NoStop}%
\bibitem [{\citenamefont {Brunton}\ \emph {et~al.}(2017)\citenamefont {Brunton}, \citenamefont {Brunton}, \citenamefont {Proctor}, \citenamefont {Kaiser},\ and\ \citenamefont {Kutz}}]{Brunton2017-xi}%
  \BibitemOpen
  \bibfield  {author} {\bibinfo {author} {\bibfnamefont {S.~L.}\ \bibnamefont {Brunton}}, \bibinfo {author} {\bibfnamefont {B.~W.}\ \bibnamefont {Brunton}}, \bibinfo {author} {\bibfnamefont {J.~L.}\ \bibnamefont {Proctor}}, \bibinfo {author} {\bibfnamefont {E.}~\bibnamefont {Kaiser}},\ and\ \bibinfo {author} {\bibfnamefont {J.~N.}\ \bibnamefont {Kutz}},\ }\bibfield  {title} {{\selectlanguage {en}\enquote {\bibinfo {title} {Chaos as an intermittently forced linear system},}\ }}\href@noop {} {\bibfield  {journal} {\bibinfo  {journal} {Nat. Commun.}\ }\textbf {\bibinfo {volume} {8}},\ \bibinfo {pages} {19} (\bibinfo {year} {2017})}\BibitemShut {NoStop}%
\bibitem [{\citenamefont {Sugihara}\ \emph {et~al.}(2012)\citenamefont {Sugihara}, \citenamefont {May}, \citenamefont {Ye}, \citenamefont {Hsieh}, \citenamefont {Deyle}, \citenamefont {Fogarty},\ and\ \citenamefont {Munch}}]{Sugihara2012-in}%
  \BibitemOpen
  \bibfield  {author} {\bibinfo {author} {\bibfnamefont {G.}~\bibnamefont {Sugihara}}, \bibinfo {author} {\bibfnamefont {R.}~\bibnamefont {May}}, \bibinfo {author} {\bibfnamefont {H.}~\bibnamefont {Ye}}, \bibinfo {author} {\bibfnamefont {C.-H.}\ \bibnamefont {Hsieh}}, \bibinfo {author} {\bibfnamefont {E.}~\bibnamefont {Deyle}}, \bibinfo {author} {\bibfnamefont {M.}~\bibnamefont {Fogarty}},\ and\ \bibinfo {author} {\bibfnamefont {S.}~\bibnamefont {Munch}},\ }\bibfield  {title} {{\selectlanguage {en}\enquote {\bibinfo {title} {Detecting causality in complex ecosystems},}\ }}\href@noop {} {\bibfield  {journal} {\bibinfo  {journal} {Science}\ }\textbf {\bibinfo {volume} {338}},\ \bibinfo {pages} {496--500} (\bibinfo {year} {2012})}\BibitemShut {NoStop}%
\bibitem [{\citenamefont {Perea}\ and\ \citenamefont {Harer}(2015)}]{Perea2015-xp}%
  \BibitemOpen
  \bibfield  {author} {\bibinfo {author} {\bibfnamefont {J.~A.}\ \bibnamefont {Perea}}\ and\ \bibinfo {author} {\bibfnamefont {J.}~\bibnamefont {Harer}},\ }\bibfield  {title} {{\selectlanguage {en}\enquote {\bibinfo {title} {Sliding windows and persistence: An application of topological methods to signal analysis},}\ }}\href@noop {} {\bibfield  {journal} {\bibinfo  {journal} {Found. Comut. Math.}\ }\textbf {\bibinfo {volume} {15}},\ \bibinfo {pages} {799--838} (\bibinfo {year} {2015})}\BibitemShut {NoStop}%
\bibitem [{\citenamefont {Williams}, \citenamefont {Kevrekidis},\ and\ \citenamefont {Rowley}(2015)}]{Williams2015-jg}%
  \BibitemOpen
  \bibfield  {author} {\bibinfo {author} {\bibfnamefont {M.~O.}\ \bibnamefont {Williams}}, \bibinfo {author} {\bibfnamefont {I.~G.}\ \bibnamefont {Kevrekidis}},\ and\ \bibinfo {author} {\bibfnamefont {C.~W.}\ \bibnamefont {Rowley}},\ }\bibfield  {title} {{\selectlanguage {en}\enquote {\bibinfo {title} {A data–driven approximation of the koopman operator: Extending dynamic mode decomposition},}\ }}\href@noop {} {\bibfield  {journal} {\bibinfo  {journal} {J. Nonlinear Sci.}\ }\textbf {\bibinfo {volume} {25}},\ \bibinfo {pages} {1307--1346} (\bibinfo {year} {2015})}\BibitemShut {NoStop}%
\bibitem [{\citenamefont {Brunton}, \citenamefont {Proctor},\ and\ \citenamefont {Kutz}(2016)}]{Brunton2016-gc}%
  \BibitemOpen
  \bibfield  {author} {\bibinfo {author} {\bibfnamefont {S.~L.}\ \bibnamefont {Brunton}}, \bibinfo {author} {\bibfnamefont {J.~L.}\ \bibnamefont {Proctor}},\ and\ \bibinfo {author} {\bibfnamefont {J.~N.}\ \bibnamefont {Kutz}},\ }\bibfield  {title} {{\selectlanguage {en}\enquote {\bibinfo {title} {Discovering governing equations from data by sparse identification of nonlinear dynamical systems},}\ }}\href@noop {} {\bibfield  {journal} {\bibinfo  {journal} {Proc. Natl. Acad. Sci. U. S. A.}\ }\textbf {\bibinfo {volume} {113}},\ \bibinfo {pages} {3932--3937} (\bibinfo {year} {2016})}\BibitemShut {NoStop}%
\bibitem [{\citenamefont {Gauthier}\ \emph {et~al.}(2021)\citenamefont {Gauthier}, \citenamefont {Bollt}, \citenamefont {Griffith},\ and\ \citenamefont {Barbosa}}]{Gauthier2021-pz}%
  \BibitemOpen
  \bibfield  {author} {\bibinfo {author} {\bibfnamefont {D.~J.}\ \bibnamefont {Gauthier}}, \bibinfo {author} {\bibfnamefont {E.}~\bibnamefont {Bollt}}, \bibinfo {author} {\bibfnamefont {A.}~\bibnamefont {Griffith}},\ and\ \bibinfo {author} {\bibfnamefont {W.~A.~S.}\ \bibnamefont {Barbosa}},\ }\bibfield  {title} {{\selectlanguage {en}\enquote {\bibinfo {title} {Next generation reservoir computing},}\ }}\href@noop {} {\bibfield  {journal} {\bibinfo  {journal} {Nat. Commun.}\ }\textbf {\bibinfo {volume} {12}},\ \bibinfo {pages} {5564} (\bibinfo {year} {2021})}\BibitemShut {NoStop}%
\bibitem [{\citenamefont {Bollt}(2021)}]{Bollt2021-mj}%
  \BibitemOpen
  \bibfield  {author} {\bibinfo {author} {\bibfnamefont {E.}~\bibnamefont {Bollt}},\ }\bibfield  {title} {{\selectlanguage {en}\enquote {\bibinfo {title} {On explaining the surprising success of reservoir computing forecaster of chaos? the universal machine learning dynamical system with contrast to {VAR} and {DMD}},}\ }}\href@noop {} {\bibfield  {journal} {\bibinfo  {journal} {Chaos}\ }\textbf {\bibinfo {volume} {31}},\ \bibinfo {pages} {013108} (\bibinfo {year} {2021})}\BibitemShut {NoStop}%
\bibitem [{\citenamefont {Schölkopf}\ and\ \citenamefont {Smola}(2002)}]{Scholkopf2002-wd}%
  \BibitemOpen
  \bibfield  {author} {\bibinfo {author} {\bibfnamefont {B.}~\bibnamefont {Schölkopf}}\ and\ \bibinfo {author} {\bibfnamefont {A.~J.}\ \bibnamefont {Smola}},\ }\href@noop {} {\emph {\bibinfo {title} {Learning with kernels: support vector machines, regularization, optimization, and beyond}}}\ (\bibinfo  {publisher} {MIT press},\ \bibinfo {year} {2002})\BibitemShut {NoStop}%
\bibitem [{\citenamefont {Kawahara}(2016)}]{Kawahara2016-ai}%
  \BibitemOpen
  \bibfield  {author} {\bibinfo {author} {\bibfnamefont {Y.}~\bibnamefont {Kawahara}},\ }\bibfield  {title} {\enquote {\bibinfo {title} {Dynamic mode decomposition with reproducing kernels for koopman spectral analysis},}\ }in\ \href@noop {} {\emph {\bibinfo {booktitle} {Advances in Neural Information Processing Systems}}},\ Vol.~\bibinfo {volume} {29}\ (\bibinfo {year} {2016})\BibitemShut {NoStop}%
\bibitem [{\citenamefont {Gu}, \citenamefont {Goel},\ and\ \citenamefont {Ré}(2021)}]{Gu2021-uh}%
  \BibitemOpen
  \bibfield  {author} {\bibinfo {author} {\bibfnamefont {A.}~\bibnamefont {Gu}}, \bibinfo {author} {\bibfnamefont {K.}~\bibnamefont {Goel}},\ and\ \bibinfo {author} {\bibfnamefont {C.}~\bibnamefont {Ré}},\ }\bibfield  {title} {\enquote {\bibinfo {title} {Efficiently modeling long sequences with structured state spaces},}\ }\href@noop {} {\bibfield  {journal} {\bibinfo  {journal} {arXiv [cs.LG]}\ } (\bibinfo {year} {2021})}\BibitemShut {NoStop}%
\bibitem [{\citenamefont {Gu}\ and\ \citenamefont {Dao}(2023)}]{Gu2023-xy}%
  \BibitemOpen
  \bibfield  {author} {\bibinfo {author} {\bibfnamefont {A.}~\bibnamefont {Gu}}\ and\ \bibinfo {author} {\bibfnamefont {T.}~\bibnamefont {Dao}},\ }\bibfield  {title} {\enquote {\bibinfo {title} {Mamba: Linear-time sequence modeling with selective state spaces},}\ }\href@noop {} {\bibfield  {journal} {\bibinfo  {journal} {arXiv [cs.LG]}\ } (\bibinfo {year} {2023})}\BibitemShut {NoStop}%
\bibitem [{\citenamefont {Lusch}, \citenamefont {Kutz},\ and\ \citenamefont {Brunton}(2018)}]{Lusch2018-yn}%
  \BibitemOpen
  \bibfield  {author} {\bibinfo {author} {\bibfnamefont {B.}~\bibnamefont {Lusch}}, \bibinfo {author} {\bibfnamefont {J.~N.}\ \bibnamefont {Kutz}},\ and\ \bibinfo {author} {\bibfnamefont {S.~L.}\ \bibnamefont {Brunton}},\ }\bibfield  {title} {{\selectlanguage {en}\enquote {\bibinfo {title} {Deep learning for universal linear embeddings of nonlinear dynamics},}\ }}\href@noop {} {\bibfield  {journal} {\bibinfo  {journal} {Nat. Commun.}\ }\textbf {\bibinfo {volume} {9}},\ \bibinfo {pages} {4950} (\bibinfo {year} {2018})}\BibitemShut {NoStop}%
\bibitem [{\citenamefont {Champion}\ \emph {et~al.}(2019)\citenamefont {Champion}, \citenamefont {Lusch}, \citenamefont {Kutz},\ and\ \citenamefont {Brunton}}]{Champion2019-nc}%
  \BibitemOpen
  \bibfield  {author} {\bibinfo {author} {\bibfnamefont {K.}~\bibnamefont {Champion}}, \bibinfo {author} {\bibfnamefont {B.}~\bibnamefont {Lusch}}, \bibinfo {author} {\bibfnamefont {J.~N.}\ \bibnamefont {Kutz}},\ and\ \bibinfo {author} {\bibfnamefont {S.~L.}\ \bibnamefont {Brunton}},\ }\bibfield  {title} {{\selectlanguage {en}\enquote {\bibinfo {title} {Data-driven discovery of coordinates and governing equations},}\ }}\href@noop {} {\bibfield  {journal} {\bibinfo  {journal} {Proc. Natl. Acad. Sci. U. S. A.}\ }\textbf {\bibinfo {volume} {116}},\ \bibinfo {pages} {22445--22451} (\bibinfo {year} {2019})}\BibitemShut {NoStop}%
\bibitem [{\citenamefont {Bakarji}\ \emph {et~al.}(2023)\citenamefont {Bakarji}, \citenamefont {Champion}, \citenamefont {Nathan~Kutz},\ and\ \citenamefont {Brunton}}]{Bakarji2023-qa}%
  \BibitemOpen
  \bibfield  {author} {\bibinfo {author} {\bibfnamefont {J.}~\bibnamefont {Bakarji}}, \bibinfo {author} {\bibfnamefont {K.}~\bibnamefont {Champion}}, \bibinfo {author} {\bibfnamefont {J.}~\bibnamefont {Nathan~Kutz}},\ and\ \bibinfo {author} {\bibfnamefont {S.~L.}\ \bibnamefont {Brunton}},\ }\bibfield  {title} {{\selectlanguage {en}\enquote {\bibinfo {title} {Discovering governing equations from partial measurements with deep delay autoencoders},}\ }}\href@noop {} {\bibfield  {journal} {\bibinfo  {journal} {Proc. Math. Phys. Eng. Sci.}\ }\textbf {\bibinfo {volume} {479}} (\bibinfo {year} {2023})}\BibitemShut {NoStop}%
\bibitem [{\citenamefont {Jaeger}(2001)}]{Jaeger2001-vs}%
  \BibitemOpen
  \bibfield  {author} {\bibinfo {author} {\bibfnamefont {H.}~\bibnamefont {Jaeger}},\ }\bibfield  {title} {\enquote {\bibinfo {title} {The ``echo state'' approach to analysing and training recurrent neural networks},}\ }\href@noop {} {\bibfield  {journal} {\bibinfo  {journal} {GMD Technical Report}\ } (\bibinfo {year} {2001})}\BibitemShut {NoStop}%
\bibitem [{\citenamefont {Maass}, \citenamefont {Natschläger},\ and\ \citenamefont {Markram}(2002)}]{Maass2002-me}%
  \BibitemOpen
  \bibfield  {author} {\bibinfo {author} {\bibfnamefont {W.}~\bibnamefont {Maass}}, \bibinfo {author} {\bibfnamefont {T.}~\bibnamefont {Natschläger}},\ and\ \bibinfo {author} {\bibfnamefont {H.}~\bibnamefont {Markram}},\ }\bibfield  {title} {{\selectlanguage {en}\enquote {\bibinfo {title} {Real-time computing without stable states: a new framework for neural computation based on perturbations},}\ }}\href@noop {} {\bibfield  {journal} {\bibinfo  {journal} {Neural Comput.}\ }\textbf {\bibinfo {volume} {14}},\ \bibinfo {pages} {2531--2560} (\bibinfo {year} {2002})}\BibitemShut {NoStop}%
\bibitem [{\citenamefont {Jaeger}(2007)}]{Jaeger2007-py}%
  \BibitemOpen
  \bibfield  {author} {\bibinfo {author} {\bibfnamefont {H.}~\bibnamefont {Jaeger}},\ }\bibfield  {title} {\enquote {\bibinfo {title} {Echo state network},}\ }\href@noop {} {\bibfield  {journal} {\bibinfo  {journal} {Scholarpedia}\ }\textbf {\bibinfo {volume} {2}},\ \bibinfo {pages} {2330} (\bibinfo {year} {2007})}\BibitemShut {NoStop}%
\bibitem [{\citenamefont {Tanaka}\ \emph {et~al.}(2019)\citenamefont {Tanaka}, \citenamefont {Yamane}, \citenamefont {Héroux}, \citenamefont {Nakane}, \citenamefont {Kanazawa}, \citenamefont {Takeda}, \citenamefont {Numata}, \citenamefont {Nakano},\ and\ \citenamefont {Hirose}}]{Tanaka2019-vn}%
  \BibitemOpen
  \bibfield  {author} {\bibinfo {author} {\bibfnamefont {G.}~\bibnamefont {Tanaka}}, \bibinfo {author} {\bibfnamefont {T.}~\bibnamefont {Yamane}}, \bibinfo {author} {\bibfnamefont {J.~B.}\ \bibnamefont {Héroux}}, \bibinfo {author} {\bibfnamefont {R.}~\bibnamefont {Nakane}}, \bibinfo {author} {\bibfnamefont {N.}~\bibnamefont {Kanazawa}}, \bibinfo {author} {\bibfnamefont {S.}~\bibnamefont {Takeda}}, \bibinfo {author} {\bibfnamefont {H.}~\bibnamefont {Numata}}, \bibinfo {author} {\bibfnamefont {D.}~\bibnamefont {Nakano}},\ and\ \bibinfo {author} {\bibfnamefont {A.}~\bibnamefont {Hirose}},\ }\bibfield  {title} {{\selectlanguage {en}\enquote {\bibinfo {title} {Recent advances in physical reservoir computing: A review},}\ }}\href@noop {} {\bibfield  {journal} {\bibinfo  {journal} {Neural Netw.}\ }\textbf {\bibinfo {volume} {115}},\ \bibinfo {pages} {100--123} (\bibinfo {year} {2019})}\BibitemShut {NoStop}%
\bibitem [{\citenamefont {Pathak}\ \emph {et~al.}(2017)\citenamefont {Pathak}, \citenamefont {Lu}, \citenamefont {Hunt}, \citenamefont {Girvan},\ and\ \citenamefont {Ott}}]{Pathak2017-id}%
  \BibitemOpen
  \bibfield  {author} {\bibinfo {author} {\bibfnamefont {J.}~\bibnamefont {Pathak}}, \bibinfo {author} {\bibfnamefont {Z.}~\bibnamefont {Lu}}, \bibinfo {author} {\bibfnamefont {B.~R.}\ \bibnamefont {Hunt}}, \bibinfo {author} {\bibfnamefont {M.}~\bibnamefont {Girvan}},\ and\ \bibinfo {author} {\bibfnamefont {E.}~\bibnamefont {Ott}},\ }\bibfield  {title} {{\selectlanguage {en}\enquote {\bibinfo {title} {Using machine learning to replicate chaotic attractors and calculate lyapunov exponents from data},}\ }}\href@noop {} {\bibfield  {journal} {\bibinfo  {journal} {Chaos}\ }\textbf {\bibinfo {volume} {27}},\ \bibinfo {pages} {121102} (\bibinfo {year} {2017})}\BibitemShut {NoStop}%
\bibitem [{\citenamefont {Lu}, \citenamefont {Hunt},\ and\ \citenamefont {Ott}(2018)}]{Lu2018-vk}%
  \BibitemOpen
  \bibfield  {author} {\bibinfo {author} {\bibfnamefont {Z.}~\bibnamefont {Lu}}, \bibinfo {author} {\bibfnamefont {B.~R.}\ \bibnamefont {Hunt}},\ and\ \bibinfo {author} {\bibfnamefont {E.}~\bibnamefont {Ott}},\ }\bibfield  {title} {{\selectlanguage {en}\enquote {\bibinfo {title} {Attractor reconstruction by machine learning},}\ }}\href@noop {} {\bibfield  {journal} {\bibinfo  {journal} {Chaos}\ }\textbf {\bibinfo {volume} {28}},\ \bibinfo {pages} {061104} (\bibinfo {year} {2018})}\BibitemShut {NoStop}%
\bibitem [{\citenamefont {Grigoryeva}, \citenamefont {Hart},\ and\ \citenamefont {Ortega}(2021)}]{Grigoryeva2021-mg}%
  \BibitemOpen
  \bibfield  {author} {\bibinfo {author} {\bibfnamefont {L.}~\bibnamefont {Grigoryeva}}, \bibinfo {author} {\bibfnamefont {A.}~\bibnamefont {Hart}},\ and\ \bibinfo {author} {\bibfnamefont {J.-P.}\ \bibnamefont {Ortega}},\ }\bibfield  {title} {{\selectlanguage {en}\enquote {\bibinfo {title} {Chaos on compact manifolds: Differentiable synchronizations beyond the takens theorem},}\ }}\href@noop {} {\bibfield  {journal} {\bibinfo  {journal} {Phys Rev E}\ }\textbf {\bibinfo {volume} {103}},\ \bibinfo {pages} {062204} (\bibinfo {year} {2021})}\BibitemShut {NoStop}%
\bibitem [{\citenamefont {Oishi}\ \emph {et~al.}(2026)\citenamefont {Oishi}, \citenamefont {Yamashita}, \citenamefont {Suzuki},\ and\ \citenamefont {Shirasaka}}]{Oishi2026-fz}%
  \BibitemOpen
  \bibfield  {author} {\bibinfo {author} {\bibfnamefont {S.}~\bibnamefont {Oishi}}, \bibinfo {author} {\bibfnamefont {H.}~\bibnamefont {Yamashita}}, \bibinfo {author} {\bibfnamefont {H.}~\bibnamefont {Suzuki}},\ and\ \bibinfo {author} {\bibfnamefont {S.}~\bibnamefont {Shirasaka}},\ }\bibfield  {title} {\enquote {\bibinfo {title} {Stabilizing chaotic dynamical system reproduction in reservoir computing},}\ }\href@noop {} {\bibfield  {journal} {\bibinfo  {journal} {arXiv [nlin.CD]}\ } (\bibinfo {year} {2026})}\BibitemShut {NoStop}%
\bibitem [{\citenamefont {Platt}\ \emph {et~al.}(2022)\citenamefont {Platt}, \citenamefont {Penny}, \citenamefont {Smith}, \citenamefont {Chen},\ and\ \citenamefont {Abarbanel}}]{Platt2022-es}%
  \BibitemOpen
  \bibfield  {author} {\bibinfo {author} {\bibfnamefont {J.~A.}\ \bibnamefont {Platt}}, \bibinfo {author} {\bibfnamefont {S.~G.}\ \bibnamefont {Penny}}, \bibinfo {author} {\bibfnamefont {T.~A.}\ \bibnamefont {Smith}}, \bibinfo {author} {\bibfnamefont {T.-C.}\ \bibnamefont {Chen}},\ and\ \bibinfo {author} {\bibfnamefont {H.~D.~I.}\ \bibnamefont {Abarbanel}},\ }\bibfield  {title} {{\selectlanguage {en}\enquote {\bibinfo {title} {A systematic exploration of reservoir computing for forecasting complex spatiotemporal dynamics},}\ }}\href@noop {} {\bibfield  {journal} {\bibinfo  {journal} {Neural Netw.}\ }\textbf {\bibinfo {volume} {153}},\ \bibinfo {pages} {530--552} (\bibinfo {year} {2022})}\BibitemShut {NoStop}%
\bibitem [{\citenamefont {Hart}(2024)}]{Hart2024-jx}%
  \BibitemOpen
  \bibfield  {author} {\bibinfo {author} {\bibfnamefont {J.~D.}\ \bibnamefont {Hart}},\ }\bibfield  {title} {{\selectlanguage {en}\enquote {\bibinfo {title} {Attractor reconstruction with reservoir computers: The effect of the reservoir's conditional lyapunov exponents on faithful attractor reconstruction},}\ }}\href@noop {} {\bibfield  {journal} {\bibinfo  {journal} {Chaos}\ }\textbf {\bibinfo {volume} {34}},\ \bibinfo {pages} {043123} (\bibinfo {year} {2024})}\BibitemShut {NoStop}%
\bibitem [{\citenamefont {Yperman}\ and\ \citenamefont {Becker}(2016)}]{Yperman2016-xd}%
  \BibitemOpen
  \bibfield  {author} {\bibinfo {author} {\bibfnamefont {J.}~\bibnamefont {Yperman}}\ and\ \bibinfo {author} {\bibfnamefont {T.}~\bibnamefont {Becker}},\ }\bibfield  {title} {\enquote {\bibinfo {title} {Bayesian optimization of hyper-parameters in reservoir computing},}\ }\href@noop {} {\bibfield  {journal} {\bibinfo  {journal} {arXiv [cs.LG]}\ } (\bibinfo {year} {2016})}\BibitemShut {NoStop}%
\bibitem [{\citenamefont {Bala}\ \emph {et~al.}(2018)\citenamefont {Bala}, \citenamefont {Ismail}, \citenamefont {Ibrahim},\ and\ \citenamefont {Sait}}]{Bala2018-zq}%
  \BibitemOpen
  \bibfield  {author} {\bibinfo {author} {\bibfnamefont {A.}~\bibnamefont {Bala}}, \bibinfo {author} {\bibfnamefont {I.}~\bibnamefont {Ismail}}, \bibinfo {author} {\bibfnamefont {R.}~\bibnamefont {Ibrahim}},\ and\ \bibinfo {author} {\bibfnamefont {S.~M.}\ \bibnamefont {Sait}},\ }\bibfield  {title} {\enquote {\bibinfo {title} {Applications of metaheuristics in reservoir computing techniques: A review},}\ }\href@noop {} {\bibfield  {journal} {\bibinfo  {journal} {IEEE Access}\ }\textbf {\bibinfo {volume} {6}},\ \bibinfo {pages} {58012--58029} (\bibinfo {year} {2018})}\BibitemShut {NoStop}%
\bibitem [{\citenamefont {Racca}\ and\ \citenamefont {Magri}(2021)}]{Racca2021-xx}%
  \BibitemOpen
  \bibfield  {author} {\bibinfo {author} {\bibfnamefont {A.}~\bibnamefont {Racca}}\ and\ \bibinfo {author} {\bibfnamefont {L.}~\bibnamefont {Magri}},\ }\bibfield  {title} {{\selectlanguage {en}\enquote {\bibinfo {title} {Robust optimization and validation of echo state networks for learning chaotic dynamics},}\ }}\href@noop {} {\bibfield  {journal} {\bibinfo  {journal} {Neural Netw.}\ }\textbf {\bibinfo {volume} {142}},\ \bibinfo {pages} {252--268} (\bibinfo {year} {2021})}\BibitemShut {NoStop}%
\bibitem [{\citenamefont {Platt}\ \emph {et~al.}(2023)\citenamefont {Platt}, \citenamefont {Penny}, \citenamefont {Smith}, \citenamefont {Chen},\ and\ \citenamefont {Abarbanel}}]{Platt2023-lh}%
  \BibitemOpen
  \bibfield  {author} {\bibinfo {author} {\bibfnamefont {J.~A.}\ \bibnamefont {Platt}}, \bibinfo {author} {\bibfnamefont {S.~G.}\ \bibnamefont {Penny}}, \bibinfo {author} {\bibfnamefont {T.~A.}\ \bibnamefont {Smith}}, \bibinfo {author} {\bibfnamefont {T.-C.}\ \bibnamefont {Chen}},\ and\ \bibinfo {author} {\bibfnamefont {H.~D.~I.}\ \bibnamefont {Abarbanel}},\ }\bibfield  {title} {{\selectlanguage {en}\enquote {\bibinfo {title} {Constraining chaos: Enforcing dynamical invariants in the training of reservoir computers},}\ }}\href@noop {} {\bibfield  {journal} {\bibinfo  {journal} {Chaos}\ }\textbf {\bibinfo {volume} {33}},\ \bibinfo {pages} {103107} (\bibinfo {year} {2023})}\BibitemShut {NoStop}%
\bibitem [{\citenamefont {Mahata}, \citenamefont {Padhi},\ and\ \citenamefont {Apte}(2023)}]{Mahata2023-dw}%
  \BibitemOpen
  \bibfield  {author} {\bibinfo {author} {\bibfnamefont {A.}~\bibnamefont {Mahata}}, \bibinfo {author} {\bibfnamefont {R.}~\bibnamefont {Padhi}},\ and\ \bibinfo {author} {\bibfnamefont {A.}~\bibnamefont {Apte}},\ }\bibfield  {title} {{\selectlanguage {en}\enquote {\bibinfo {title} {Variability of echo state network prediction horizon for partially observed dynamical systems},}\ }}\href@noop {} {\bibfield  {journal} {\bibinfo  {journal} {Phys. Rev. E.}\ }\textbf {\bibinfo {volume} {108}},\ \bibinfo {pages} {064209} (\bibinfo {year} {2023})}\BibitemShut {NoStop}%
\bibitem [{\citenamefont {Haller}\ and\ \citenamefont {Ponsioen}(2016)}]{Haller2016-oj}%
  \BibitemOpen
  \bibfield  {author} {\bibinfo {author} {\bibfnamefont {G.}~\bibnamefont {Haller}}\ and\ \bibinfo {author} {\bibfnamefont {S.}~\bibnamefont {Ponsioen}},\ }\bibfield  {title} {{\selectlanguage {en}\enquote {\bibinfo {title} {Nonlinear normal modes and spectral submanifolds: existence, uniqueness and use in model reduction},}\ }}\href@noop {} {\bibfield  {journal} {\bibinfo  {journal} {Nonlinear Dyn.}\ }\textbf {\bibinfo {volume} {86}},\ \bibinfo {pages} {1493--1534} (\bibinfo {year} {2016})}\BibitemShut {NoStop}%
\bibitem [{\citenamefont {Jaeger}\ \emph {et~al.}(2007)\citenamefont {Jaeger}, \citenamefont {Lukosevicius}, \citenamefont {Popovici},\ and\ \citenamefont {Siewert}}]{Jaeger2007-kc}%
  \BibitemOpen
  \bibfield  {author} {\bibinfo {author} {\bibfnamefont {H.}~\bibnamefont {Jaeger}}, \bibinfo {author} {\bibfnamefont {M.}~\bibnamefont {Lukosevicius}}, \bibinfo {author} {\bibfnamefont {D.}~\bibnamefont {Popovici}},\ and\ \bibinfo {author} {\bibfnamefont {U.}~\bibnamefont {Siewert}},\ }\bibfield  {title} {{\selectlanguage {en}\enquote {\bibinfo {title} {Optimization and applications of echo state networks with leaky-integrator neurons},}\ }}\href@noop {} {\bibfield  {journal} {\bibinfo  {journal} {Neural Netw.}\ }\textbf {\bibinfo {volume} {20}},\ \bibinfo {pages} {335--352} (\bibinfo {year} {2007})}\BibitemShut {NoStop}%
\bibitem [{\citenamefont {Rulkov}\ \emph {et~al.}(1995)\citenamefont {Rulkov}, \citenamefont {Sushchik}, \citenamefont {Tsimring},\ and\ \citenamefont {Abarbanel}}]{Rulkov1995-tq}%
  \BibitemOpen
  \bibfield  {author} {\bibinfo {author} {\bibfnamefont {N.~F.}\ \bibnamefont {Rulkov}}, \bibinfo {author} {\bibfnamefont {M.~M.}\ \bibnamefont {Sushchik}}, \bibinfo {author} {\bibfnamefont {L.~S.}\ \bibnamefont {Tsimring}},\ and\ \bibinfo {author} {\bibfnamefont {H.~D.}\ \bibnamefont {Abarbanel}},\ }\bibfield  {title} {{\selectlanguage {en}\enquote {\bibinfo {title} {Generalized synchronization of chaos in directionally coupled chaotic systems},}\ }}\href@noop {} {\bibfield  {journal} {\bibinfo  {journal} {Phys. Rev. E Stat. Phys. Plasmas Fluids Relat. Interdiscip. Topics}\ }\textbf {\bibinfo {volume} {51}},\ \bibinfo {pages} {980--994} (\bibinfo {year} {1995})}\BibitemShut {NoStop}%
\bibitem [{\citenamefont {Pyragas}(1996)}]{Pyragas1996-kv}%
  \BibitemOpen
  \bibfield  {author} {\bibinfo {author} {\bibfnamefont {K.}~\bibnamefont {Pyragas}},\ }\bibfield  {title} {{\selectlanguage {en}\enquote {\bibinfo {title} {Weak and strong synchronization of chaos},}\ }}\href@noop {} {\bibfield  {journal} {\bibinfo  {journal} {Phys. Rev. E Stat. Phys. Plasmas Fluids Relat. Interdiscip. Topics}\ }\textbf {\bibinfo {volume} {54}},\ \bibinfo {pages} {R4508--R4511} (\bibinfo {year} {1996})}\BibitemShut {NoStop}%
\bibitem [{\citenamefont {Suetani}\ and\ \citenamefont {Parlitz}(2026)}]{Suetani2026-hg}%
  \BibitemOpen
  \bibfield  {author} {\bibinfo {author} {\bibfnamefont {H.}~\bibnamefont {Suetani}}\ and\ \bibinfo {author} {\bibfnamefont {U.}~\bibnamefont {Parlitz}},\ }\bibfield  {title} {{\selectlanguage {en}\enquote {\bibinfo {title} {Impact of weak generalized synchronization on time series forecasting using reservoir computers},}\ }}\href@noop {} {\bibfield  {journal} {\bibinfo  {journal} {Chaos}\ }\textbf {\bibinfo {volume} {36}},\ \bibinfo {pages} {043125} (\bibinfo {year} {2026})}\BibitemShut {NoStop}%
\bibitem [{\citenamefont {Manjunath}\ and\ \citenamefont {Jaeger}(2013)}]{Manjunath2013-bj}%
  \BibitemOpen
  \bibfield  {author} {\bibinfo {author} {\bibfnamefont {G.}~\bibnamefont {Manjunath}}\ and\ \bibinfo {author} {\bibfnamefont {H.}~\bibnamefont {Jaeger}},\ }\bibfield  {title} {{\selectlanguage {en}\enquote {\bibinfo {title} {Echo state property linked to an input: exploring a fundamental characteristic of recurrent neural networks},}\ }}\href@noop {} {\bibfield  {journal} {\bibinfo  {journal} {Neural Comput.}\ }\textbf {\bibinfo {volume} {25}},\ \bibinfo {pages} {671--696} (\bibinfo {year} {2013})}\BibitemShut {NoStop}%
\bibitem [{\citenamefont {Hart}, \citenamefont {Hook},\ and\ \citenamefont {Dawes}(2020)}]{Hart2020-kx}%
  \BibitemOpen
  \bibfield  {author} {\bibinfo {author} {\bibfnamefont {A.}~\bibnamefont {Hart}}, \bibinfo {author} {\bibfnamefont {J.}~\bibnamefont {Hook}},\ and\ \bibinfo {author} {\bibfnamefont {J.}~\bibnamefont {Dawes}},\ }\bibfield  {title} {{\selectlanguage {en}\enquote {\bibinfo {title} {Embedding and approximation theorems for echo state networks},}\ }}\href@noop {} {\bibfield  {journal} {\bibinfo  {journal} {Neural Netw.}\ }\textbf {\bibinfo {volume} {128}},\ \bibinfo {pages} {234--247} (\bibinfo {year} {2020})}\BibitemShut {NoStop}%
\bibitem [{\citenamefont {Grigoryeva}, \citenamefont {Hart},\ and\ \citenamefont {Ortega}(2023)}]{Grigoryeva2023-cw}%
  \BibitemOpen
  \bibfield  {author} {\bibinfo {author} {\bibfnamefont {L.}~\bibnamefont {Grigoryeva}}, \bibinfo {author} {\bibfnamefont {A.}~\bibnamefont {Hart}},\ and\ \bibinfo {author} {\bibfnamefont {J.-P.}\ \bibnamefont {Ortega}},\ }\bibfield  {title} {{\selectlanguage {en}\enquote {\bibinfo {title} {Learning strange attractors with reservoir systems},}\ }}\href@noop {} {\bibfield  {journal} {\bibinfo  {journal} {Nonlinearity}\ }\textbf {\bibinfo {volume} {36}},\ \bibinfo {pages} {4674} (\bibinfo {year} {2023})}\BibitemShut {NoStop}%
\bibitem [{\citenamefont {Bishop}(2006)}]{Bishop2006-iz}%
  \BibitemOpen
  \bibfield  {author} {\bibinfo {author} {\bibfnamefont {C.~M.}\ \bibnamefont {Bishop}},\ }\href@noop {} {\emph {\bibinfo {title} {Pattern Recognition and Machine Learning}}}\ (\bibinfo  {publisher} {Springer New York},\ \bibinfo {year} {2006})\BibitemShut {NoStop}%
\bibitem [{\citenamefont {Fujisaka}\ and\ \citenamefont {Yamada}(1983)}]{Fujisaka1983-hu}%
  \BibitemOpen
  \bibfield  {author} {\bibinfo {author} {\bibfnamefont {H.}~\bibnamefont {Fujisaka}}\ and\ \bibinfo {author} {\bibfnamefont {T.}~\bibnamefont {Yamada}},\ }\bibfield  {title} {\enquote {\bibinfo {title} {Stability theory of synchronized motion in coupled-oscillator systems:},}\ }\href@noop {} {\bibfield  {journal} {\bibinfo  {journal} {Prog Theor Phys}\ }\textbf {\bibinfo {volume} {69}},\ \bibinfo {pages} {32--47} (\bibinfo {year} {1983})}\BibitemShut {NoStop}%
\bibitem [{\citenamefont {Pecora}\ and\ \citenamefont {Carroll}(2015)}]{Pecora2015-gy}%
  \BibitemOpen
  \bibfield  {author} {\bibinfo {author} {\bibfnamefont {L.~M.}\ \bibnamefont {Pecora}}\ and\ \bibinfo {author} {\bibfnamefont {T.~L.}\ \bibnamefont {Carroll}},\ }\bibfield  {title} {{\selectlanguage {en}\enquote {\bibinfo {title} {Synchronization of chaotic systems},}\ }}\href@noop {} {\bibfield  {journal} {\bibinfo  {journal} {Chaos}\ }\textbf {\bibinfo {volume} {25}},\ \bibinfo {pages} {097611} (\bibinfo {year} {2015})}\BibitemShut {NoStop}%
\bibitem [{Note1()}]{Note1}%
  \BibitemOpen
  \bibinfo {note} {The final equality follows from Eq.~\protect \eqref {eq:closed-loop-dynamics}, which allows the $\protect \bm {\tanh }$ term to be expressed directly in terms of the stored state sequence. This form is convenient for numerical computation of the Lyapunov exponents.}\BibitemShut {Stop}%
\bibitem [{\citenamefont {Lorenz}(1963)}]{Lorenz1963-kf}%
  \BibitemOpen
  \bibfield  {author} {\bibinfo {author} {\bibfnamefont {E.~N.}\ \bibnamefont {Lorenz}},\ }\bibfield  {title} {{\selectlanguage {en}\enquote {\bibinfo {title} {Deterministic nonperiodic flow},}\ }}\href@noop {} {\bibfield  {journal} {\bibinfo  {journal} {J. Atmos. Sci.}\ }\textbf {\bibinfo {volume} {20}},\ \bibinfo {pages} {130--141} (\bibinfo {year} {1963})}\BibitemShut {NoStop}%
\bibitem [{\citenamefont {Vlachas}\ \emph {et~al.}(2020)\citenamefont {Vlachas}, \citenamefont {Pathak}, \citenamefont {Hunt}, \citenamefont {Sapsis}, \citenamefont {Girvan}, \citenamefont {Ott},\ and\ \citenamefont {Koumoutsakos}}]{Vlachas2020-ob}%
  \BibitemOpen
  \bibfield  {author} {\bibinfo {author} {\bibfnamefont {P.~R.}\ \bibnamefont {Vlachas}}, \bibinfo {author} {\bibfnamefont {J.}~\bibnamefont {Pathak}}, \bibinfo {author} {\bibfnamefont {B.~R.}\ \bibnamefont {Hunt}}, \bibinfo {author} {\bibfnamefont {T.~P.}\ \bibnamefont {Sapsis}}, \bibinfo {author} {\bibfnamefont {M.}~\bibnamefont {Girvan}}, \bibinfo {author} {\bibfnamefont {E.}~\bibnamefont {Ott}},\ and\ \bibinfo {author} {\bibfnamefont {P.}~\bibnamefont {Koumoutsakos}},\ }\bibfield  {title} {{\selectlanguage {en}\enquote {\bibinfo {title} {Backpropagation algorithms and reservoir computing in recurrent neural networks for the forecasting of complex spatiotemporal dynamics},}\ }}\href@noop {} {\bibfield  {journal} {\bibinfo  {journal} {Neural Netw.}\ }\textbf {\bibinfo {volume} {126}},\ \bibinfo {pages} {191--217} (\bibinfo {year} {2020})}\BibitemShut {NoStop}%
\bibitem [{\citenamefont {Kuehn}(2015)}]{Kuehn2015-pl}%
  \BibitemOpen
  \bibfield  {author} {\bibinfo {author} {\bibfnamefont {C.}~\bibnamefont {Kuehn}},\ }\href@noop {} {{\selectlanguage {en}\emph {\bibinfo {title} {Multiple time scale dynamics}}}},\ \bibinfo {edition} {2015th}\ ed.,\ Applied Mathematical Sciences\ (\bibinfo  {publisher} {Springer International Publishing},\ \bibinfo {address} {Cham, Switzerland},\ \bibinfo {year} {2015})\BibitemShut {NoStop}%
\bibitem [{\citenamefont {Fenichel}(1979)}]{Fenichel1979-ej}%
  \BibitemOpen
  \bibfield  {author} {\bibinfo {author} {\bibfnamefont {N.}~\bibnamefont {Fenichel}},\ }\bibfield  {title} {{\selectlanguage {en}\enquote {\bibinfo {title} {Geometric singular perturbation theory for ordinary differential equations},}\ }}\href@noop {} {\bibfield  {journal} {\bibinfo  {journal} {J. Differ. Equ.}\ }\textbf {\bibinfo {volume} {31}},\ \bibinfo {pages} {53--98} (\bibinfo {year} {1979})}\BibitemShut {NoStop}%
\bibitem [{\citenamefont {Hirsch}, \citenamefont {Pugh},\ and\ \citenamefont {Shub}(1977)}]{Hirsch1977-ry}%
  \BibitemOpen
  \bibfield  {author} {\bibinfo {author} {\bibfnamefont {M.~W.}\ \bibnamefont {Hirsch}}, \bibinfo {author} {\bibfnamefont {C.~C.}\ \bibnamefont {Pugh}},\ and\ \bibinfo {author} {\bibfnamefont {M.}~\bibnamefont {Shub}},\ }\href@noop {} {{\selectlanguage {en}\emph {\bibinfo {title} {Invariant Manifolds}}}},\ Lecture notes in mathematics\ (\bibinfo  {publisher} {Springer},\ \bibinfo {address} {Berlin, Germany},\ \bibinfo {year} {1977})\BibitemShut {NoStop}%
\bibitem [{Note2()}]{Note2}%
  \BibitemOpen
  \bibinfo {note} {Preserving the eigenstructure is not essential to the proposed framework. This conservative construction is adopted here to isolate the effect of slow-spectrum preshaping. In practical implementations, simpler constructions would suffice for slow-spectrum preshaping while allowing the eigenvectors to change.}\BibitemShut {Stop}%
\bibitem [{Note3()}]{Note3}%
  \BibitemOpen
  \bibinfo {note} {The slow eigenvalues are placed in $\protect \mathbf {A}'$ by Algorithm~\ref {alg:slow-spectrum-preshaping}, not directly in $\protect \mathbf {A}'_{\protect \mathrm {eff}}$. Because $\protect \mathbf {A}'_{\protect \mathrm {eff}} = (1-\alpha )\protect \mathbf {I} + \alpha \protect \mathbf {A}'$ shares its eigenvectors with $\protect \mathbf {A}'$, an eigenvalue $\mu $ of $\protect \mathbf {A}'$ corresponds to the eigenvalue $1-\alpha +\alpha \mu $ of $\protect \mathbf {A}'_{\protect \mathrm {eff}}$. The leaking rate $\alpha $ therefore shifts the placed eigenvalues, but the associated invariant subspace is unchanged and the eigenvalues remain slow, so the goal of introducing a few slow eigenvalues into $\protect \mathbf {A}'_{\protect \mathrm {eff}}$ is still achieved.}\BibitemShut {Stop}%
\bibitem [{Note4()}]{Note4}%
  \BibitemOpen
  \bibinfo {note} {When such normalization is not applicable, an appropriate scaling factor should be selected individually according to the amplitude and range of the target signal.}\BibitemShut {Stop}%
\bibitem [{\citenamefont {Whitney}(1936)}]{Whitney1936-es}%
  \BibitemOpen
  \bibfield  {author} {\bibinfo {author} {\bibfnamefont {H.}~\bibnamefont {Whitney}},\ }\bibfield  {title} {{\selectlanguage {en}\enquote {\bibinfo {title} {Differentiable manifolds},}\ }}\href@noop {} {\bibfield  {journal} {\bibinfo  {journal} {Ann. Math.}\ }\textbf {\bibinfo {volume} {37}},\ \bibinfo {pages} {645} (\bibinfo {year} {1936})}\BibitemShut {NoStop}%
\bibitem [{\citenamefont {Kantz}\ and\ \citenamefont {Olbrich}(1997)}]{Kantz1997-lk}%
  \BibitemOpen
  \bibfield  {author} {\bibinfo {author} {\bibfnamefont {H.}~\bibnamefont {Kantz}}\ and\ \bibinfo {author} {\bibfnamefont {E.}~\bibnamefont {Olbrich}},\ }\bibfield  {title} {{\selectlanguage {en}\enquote {\bibinfo {title} {Scalar observations from a class of high-dimensional chaotic systems: Limitations of the time delay embedding},}\ }}\href@noop {} {\bibfield  {journal} {\bibinfo  {journal} {Chaos}\ }\textbf {\bibinfo {volume} {7}},\ \bibinfo {pages} {423--429} (\bibinfo {year} {1997})}\BibitemShut {NoStop}%
\bibitem [{\citenamefont {Olbrich}\ and\ \citenamefont {Kantz}(1997)}]{Olbrich1997-nx}%
  \BibitemOpen
  \bibfield  {author} {\bibinfo {author} {\bibfnamefont {E.}~\bibnamefont {Olbrich}}\ and\ \bibinfo {author} {\bibfnamefont {H.}~\bibnamefont {Kantz}},\ }\bibfield  {title} {\enquote {\bibinfo {title} {Inferring chaotic dynamics from time-series: On which length scale determinism becomes visible},}\ }\href@noop {} {\bibfield  {journal} {\bibinfo  {journal} {Phys. Lett. A}\ }\textbf {\bibinfo {volume} {232}},\ \bibinfo {pages} {63--69} (\bibinfo {year} {1997})}\BibitemShut {NoStop}%
\bibitem [{\citenamefont {Sauer}, \citenamefont {Tempkin},\ and\ \citenamefont {Yorke}(1998)}]{Sauer1998-la}%
  \BibitemOpen
  \bibfield  {author} {\bibinfo {author} {\bibfnamefont {T.~D.}\ \bibnamefont {Sauer}}, \bibinfo {author} {\bibfnamefont {J.~A.}\ \bibnamefont {Tempkin}},\ and\ \bibinfo {author} {\bibfnamefont {J.~A.}\ \bibnamefont {Yorke}},\ }\bibfield  {title} {\enquote {\bibinfo {title} {Spurious lyapunov exponents in attractor reconstruction},}\ }\href@noop {} {\bibfield  {journal} {\bibinfo  {journal} {Phys. Rev. Lett.}\ }\textbf {\bibinfo {volume} {81}},\ \bibinfo {pages} {4341--4344} (\bibinfo {year} {1998})}\BibitemShut {NoStop}%
\bibitem [{\citenamefont {Pecora}\ \emph {et~al.}(2007)\citenamefont {Pecora}, \citenamefont {Moniz}, \citenamefont {Nichols},\ and\ \citenamefont {Carroll}}]{Pecora2007-fj}%
  \BibitemOpen
  \bibfield  {author} {\bibinfo {author} {\bibfnamefont {L.~M.}\ \bibnamefont {Pecora}}, \bibinfo {author} {\bibfnamefont {L.}~\bibnamefont {Moniz}}, \bibinfo {author} {\bibfnamefont {J.}~\bibnamefont {Nichols}},\ and\ \bibinfo {author} {\bibfnamefont {T.~L.}\ \bibnamefont {Carroll}},\ }\bibfield  {title} {{\selectlanguage {en}\enquote {\bibinfo {title} {A unified approach to attractor reconstruction},}\ }}\href@noop {} {\bibfield  {journal} {\bibinfo  {journal} {Chaos}\ }\textbf {\bibinfo {volume} {17}},\ \bibinfo {pages} {013110} (\bibinfo {year} {2007})}\BibitemShut {NoStop}%
\bibitem [{\citenamefont {Pecora}\ and\ \citenamefont {Carroll}(2025)}]{Pecora2025-js}%
  \BibitemOpen
  \bibfield  {author} {\bibinfo {author} {\bibfnamefont {L.}~\bibnamefont {Pecora}}\ and\ \bibinfo {author} {\bibfnamefont {T.}~\bibnamefont {Carroll}},\ }\bibfield  {title} {{\selectlanguage {en}\enquote {\bibinfo {title} {Statistics for differential topological properties between datasets with an application to reservoir computers},}\ }}\href@noop {} {\bibfield  {journal} {\bibinfo  {journal} {Chaos}\ }\textbf {\bibinfo {volume} {35}},\ \bibinfo {pages} {073153} (\bibinfo {year} {2025})}\BibitemShut {NoStop}%
\bibitem [{\citenamefont {Kennel}, \citenamefont {Brown},\ and\ \citenamefont {Abarbanel}(1992)}]{Kennel1992-rz}%
  \BibitemOpen
  \bibfield  {author} {\bibinfo {author} {\bibfnamefont {M.~B.}\ \bibnamefont {Kennel}}, \bibinfo {author} {\bibfnamefont {R.}~\bibnamefont {Brown}},\ and\ \bibinfo {author} {\bibfnamefont {H.~D.}\ \bibnamefont {Abarbanel}},\ }\bibfield  {title} {{\selectlanguage {en}\enquote {\bibinfo {title} {Determining embedding dimension for phase-space reconstruction using a geometrical construction},}\ }}\href@noop {} {\bibfield  {journal} {\bibinfo  {journal} {Phys. Rev. A}\ }\textbf {\bibinfo {volume} {45}},\ \bibinfo {pages} {3403--3411} (\bibinfo {year} {1992})}\BibitemShut {NoStop}%
\bibitem [{\citenamefont {Liebert}, \citenamefont {Pawelzik},\ and\ \citenamefont {Schuster}(1991)}]{Liebert1991-dz}%
  \BibitemOpen
  \bibfield  {author} {\bibinfo {author} {\bibfnamefont {W.}~\bibnamefont {Liebert}}, \bibinfo {author} {\bibfnamefont {K.}~\bibnamefont {Pawelzik}},\ and\ \bibinfo {author} {\bibfnamefont {H.~G.}\ \bibnamefont {Schuster}},\ }\bibfield  {title} {\enquote {\bibinfo {title} {Optimal embeddings of chaotic attractors from topological considerations},}\ }\href@noop {} {\bibfield  {journal} {\bibinfo  {journal} {EPL}\ }\textbf {\bibinfo {volume} {14}},\ \bibinfo {pages} {521--526} (\bibinfo {year} {1991})}\BibitemShut {NoStop}%
\bibitem [{\citenamefont {Kocarev}\ and\ \citenamefont {Parlitz}(1996)}]{Kocarev1996-mt}%
  \BibitemOpen
  \bibfield  {author} {\bibinfo {author} {\bibfnamefont {L.}~\bibnamefont {Kocarev}}\ and\ \bibinfo {author} {\bibfnamefont {U.}~\bibnamefont {Parlitz}},\ }\bibfield  {title} {{\selectlanguage {en}\enquote {\bibinfo {title} {Generalized synchronization, predictability, and equivalence of unidirectionally coupled dynamical systems},}\ }}\href@noop {} {\bibfield  {journal} {\bibinfo  {journal} {Phys. Rev. Lett.}\ }\textbf {\bibinfo {volume} {76}},\ \bibinfo {pages} {1816--1819} (\bibinfo {year} {1996})}\BibitemShut {NoStop}%
\bibitem [{\citenamefont {Gao}\ and\ \citenamefont {Zheng}(1993)}]{Gao1993-fj}%
  \BibitemOpen
  \bibfield  {author} {\bibinfo {author} {\bibfnamefont {J.}~\bibnamefont {Gao}}\ and\ \bibinfo {author} {\bibfnamefont {Z.}~\bibnamefont {Zheng}},\ }\bibfield  {title} {{\selectlanguage {en}\enquote {\bibinfo {title} {Local exponential divergence plot and optimal embedding of a chaotic time series},}\ }}\href@noop {} {\bibfield  {journal} {\bibinfo  {journal} {Phys. Lett. A}\ }\textbf {\bibinfo {volume} {181}},\ \bibinfo {pages} {153--158} (\bibinfo {year} {1993})}\BibitemShut {NoStop}%
\bibitem [{\citenamefont {Gilpin}(2021)}]{Gilpin2021-wp}%
  \BibitemOpen
  \bibfield  {author} {\bibinfo {author} {\bibfnamefont {W.}~\bibnamefont {Gilpin}},\ }\bibfield  {title} {\enquote {\bibinfo {title} {Chaos as an interpretable benchmark for forecasting and data-driven modelling},}\ }\href@noop {} {\bibfield  {journal} {\bibinfo  {journal} {arXiv [cs.LG]}\ } (\bibinfo {year} {2021})}\BibitemShut {NoStop}%
\bibitem [{\citenamefont {Gilpin}(2023)}]{Gilpin2023-ua}%
  \BibitemOpen
  \bibfield  {author} {\bibinfo {author} {\bibfnamefont {W.}~\bibnamefont {Gilpin}},\ }\bibfield  {title} {{\selectlanguage {en}\enquote {\bibinfo {title} {Model scale versus domain knowledge in statistical forecasting of chaotic systems},}\ }}\href@noop {} {\bibfield  {journal} {\bibinfo  {journal} {Phys. Rev. Res.}\ }\textbf {\bibinfo {volume} {5}},\ \bibinfo {pages} {043252} (\bibinfo {year} {2023})}\BibitemShut {NoStop}%
\bibitem [{\citenamefont {Kraemer}\ \emph {et~al.}(2021)\citenamefont {Kraemer}, \citenamefont {Datseris}, \citenamefont {Kurths}, \citenamefont {Kiss}, \citenamefont {Ocampo-Espindola},\ and\ \citenamefont {Marwan}}]{Kraemer2021-ln}%
  \BibitemOpen
  \bibfield  {author} {\bibinfo {author} {\bibfnamefont {K.~H.}\ \bibnamefont {Kraemer}}, \bibinfo {author} {\bibfnamefont {G.}~\bibnamefont {Datseris}}, \bibinfo {author} {\bibfnamefont {J.}~\bibnamefont {Kurths}}, \bibinfo {author} {\bibfnamefont {I.~Z.}\ \bibnamefont {Kiss}}, \bibinfo {author} {\bibfnamefont {J.~L.}\ \bibnamefont {Ocampo-Espindola}},\ and\ \bibinfo {author} {\bibfnamefont {N.}~\bibnamefont {Marwan}},\ }\bibfield  {title} {{\selectlanguage {en}\enquote {\bibinfo {title} {A unified and automated approach to attractor reconstruction},}\ }}\href@noop {} {\bibfield  {journal} {\bibinfo  {journal} {New J. Phys.}\ }\textbf {\bibinfo {volume} {23}},\ \bibinfo {pages} {033017} (\bibinfo {year} {2021})}\BibitemShut {NoStop}%
\bibitem [{\citenamefont {Nakai}\ and\ \citenamefont {Saiki}(2018)}]{Nakai2018-cw}%
  \BibitemOpen
  \bibfield  {author} {\bibinfo {author} {\bibfnamefont {K.}~\bibnamefont {Nakai}}\ and\ \bibinfo {author} {\bibfnamefont {Y.}~\bibnamefont {Saiki}},\ }\bibfield  {title} {{\selectlanguage {en}\enquote {\bibinfo {title} {Machine-learning inference of fluid variables from data using reservoir computing},}\ }}\href@noop {} {\bibfield  {journal} {\bibinfo  {journal} {Phys. Rev. E.}\ }\textbf {\bibinfo {volume} {98}},\ \bibinfo {pages} {023111} (\bibinfo {year} {2018})}\BibitemShut {NoStop}%
\bibitem [{\citenamefont {Gottwald}\ and\ \citenamefont {Reich}(2021)}]{Gottwald2021-zx}%
  \BibitemOpen
  \bibfield  {author} {\bibinfo {author} {\bibfnamefont {G.~A.}\ \bibnamefont {Gottwald}}\ and\ \bibinfo {author} {\bibfnamefont {S.}~\bibnamefont {Reich}},\ }\bibfield  {title} {{\selectlanguage {en}\enquote {\bibinfo {title} {Combining machine learning and data assimilation to forecast dynamical systems from noisy partial observations},}\ }}\href@noop {} {\bibfield  {journal} {\bibinfo  {journal} {Chaos}\ }\textbf {\bibinfo {volume} {31}},\ \bibinfo {pages} {101103} (\bibinfo {year} {2021})}\BibitemShut {NoStop}%
\bibitem [{\citenamefont {Goswami}(2023)}]{Goswami2023-th}%
  \BibitemOpen
  \bibfield  {author} {\bibinfo {author} {\bibfnamefont {D.}~\bibnamefont {Goswami}},\ }\bibfield  {title} {{\selectlanguage {en}\enquote {\bibinfo {title} {Delay embedded echo-state network: A predictor for partially observed systems},}\ }}\href@noop {} {\bibfield  {journal} {\bibinfo  {journal} {IFAC-PapersOnLine}\ }\textbf {\bibinfo {volume} {56}},\ \bibinfo {pages} {6826--6832} (\bibinfo {year} {2023})}\BibitemShut {NoStop}%
\bibitem [{\citenamefont {Duan}\ \emph {et~al.}(2023)\citenamefont {Duan}, \citenamefont {Ying}, \citenamefont {Leng}, \citenamefont {Kurths}, \citenamefont {Lin},\ and\ \citenamefont {Ma}}]{Duan2023-ge}%
  \BibitemOpen
  \bibfield  {author} {\bibinfo {author} {\bibfnamefont {X.-Y.}\ \bibnamefont {Duan}}, \bibinfo {author} {\bibfnamefont {X.}~\bibnamefont {Ying}}, \bibinfo {author} {\bibfnamefont {S.-Y.}\ \bibnamefont {Leng}}, \bibinfo {author} {\bibfnamefont {J.}~\bibnamefont {Kurths}}, \bibinfo {author} {\bibfnamefont {W.}~\bibnamefont {Lin}},\ and\ \bibinfo {author} {\bibfnamefont {H.-F.}\ \bibnamefont {Ma}},\ }\bibfield  {title} {{\selectlanguage {en}\enquote {\bibinfo {title} {Embedding theory of reservoir computing and reducing reservoir network using time delays},}\ }}\href@noop {} {\bibfield  {journal} {\bibinfo  {journal} {Phys. Rev. Res.}\ }\textbf {\bibinfo {volume} {5}},\ \bibinfo {pages} {L022041} (\bibinfo {year} {2023})}\BibitemShut {NoStop}%
\bibitem [{\citenamefont {Jaurigue}\ \emph {et~al.}(2025)\citenamefont {Jaurigue}, \citenamefont {Robertson}, \citenamefont {Hurtado}, \citenamefont {Jaurigue},\ and\ \citenamefont {Lüdge}}]{Jaurigue2025-wy}%
  \BibitemOpen
  \bibfield  {author} {\bibinfo {author} {\bibfnamefont {J.}~\bibnamefont {Jaurigue}}, \bibinfo {author} {\bibfnamefont {J.}~\bibnamefont {Robertson}}, \bibinfo {author} {\bibfnamefont {A.}~\bibnamefont {Hurtado}}, \bibinfo {author} {\bibfnamefont {L.}~\bibnamefont {Jaurigue}},\ and\ \bibinfo {author} {\bibfnamefont {K.}~\bibnamefont {Lüdge}},\ }\bibfield  {title} {{\selectlanguage {en}\enquote {\bibinfo {title} {Post-processing methods for delay embedding and feature scaling of reservoir computers},}\ }}\href@noop {} {\bibfield  {journal} {\bibinfo  {journal} {Commun Eng}\ }\textbf {\bibinfo {volume} {4}},\ \bibinfo {pages} {10} (\bibinfo {year} {2025})}\BibitemShut {NoStop}%
\end{thebibliography}%

\end{document}